\documentclass[twocolumn,10pt]{article}
          \begin{document}
          \title{The Representation of  Numbers in Quantum
          Mechanics}

          \author{Paul Benioff\thanks{This work is supported by the U.S.
            Department of Energy, Nuclear Physics Division, under contract
            W-31-109-ENG-38.}\\
           Physics Division, Argonne National Laboratory \\
           Argonne, IL 60439 \\
           e-mail: pbenioff@anl.gov}
           \date{\today}

          \maketitle
          \begin{abstract}
          Earlier work on modular arithmetic of $k-ary$ representations of
          length $L$ of the natural numbers in quantum mechanics is extended here to
          $k-ary$ representations of all natural numbers, and to integers and rational
          numbers. Since the length $L$ is indeterminate,
          representations of states and operators using creation and annihilation
          operators for bosons and fermions are defined. Emphasis is on
          definitions and properties of operators corresponding to the basic
          operations whose properties are given by the axioms for each type of
          number. The importance of the requirement of efficient
          implementability for physical models of the axioms is
          emphasized. Based on this, successor operations for each value of
          $j$ corresponding to $+k^{j-1}$ are defined. It follows from the efficient
          implementability of these successors, which is the case for all
          computers, that implementation of the addition and multiplication operators,
          which are defined in terms of polynomially many
          iterations of the successors, should be efficient. This is
          not the case for definitions based on just the
          successor for $j=1$.  This is the only successor
          defined in the usual axioms of arithmetic.

          \end{abstract}
          \section{Introduction}
          Quantum computers are of much recent interest mainly because of their
          ability to implement some algorithms \cite{Shor,Grover}
          more efficiently than any known classical algorithms.
          Also the possibility that they can simulate other
          quantum systems more efficiently than is possible by
          classical systems \cite{Feynman} is of interest. Quantum
          robots \cite{BenQM} may also be of interest.  These are
          mobile systems including a quantum computer and
          ancillary systems that move in and interact with an
          arbitrary environment of quantum systems.

          A central aspect of computation is the fact that the physical states acted on by
          both quantum and classical computers represent numbers.
          This raises the question regarding exactly what are the
           numbers that are supposed to be
          represented by computer states.  The viewpoint usually
          taken is that one knows intuitively what numbers are and how
          to interpret the various representations. For example in
          quantum mechanics the product state
          $|\underline{s}\rangle=
          \otimes_{j=1}^{L}|\underline{s}(j)\rangle_{j}$ where $
          \underline{s}$ is a function from $1,2,\cdots ,L$ to
          $0,1$ is a binary representation of numbers according to
          \begin{equation} s=\sum_{j=1}^{L}
          \underline{s}(j)2^{j-1} \label{numrep} \end{equation}
          where the left hand symbol, $s$, denotes a natural
          number with no particular representation specified.

          Another approach is to characterize numbers as models of the axioms  for
          arithmetic or number theory \cite{Shoenfield,Smullyan}. Any mathematical
          or physical system that satisfies (or is a model of) the axioms of arithmetic or
            number theory represents the natural numbers. This
            description can be extended to other mathematical
            systems.  For example any mathematical or physical system that satisfies
          appropriate sets of axioms for integers or rational numbers is a
          representation or model of these types of
          numbers.\footnote{The intuitive base remains, though,
          as axioms are set up to reflect the intuitive properties
          of each type of number.}

          This viewpoint will be taken here as it gives a precise method for
          characterizing the various types of numbers and discussing both mathematical and
          physical models of the axioms.  This viewpoint also emphasizes the close
          connection between mathematics and physics and the
          relevance of mathematical logic to the development of a
          coherent theory of mathematics and physics. The importance of developing a
          coherent theory of mathematics and physics has been noted
          elsewhere \cite{BenDTVQM} and in other work
          \cite{Tegmark,Qsets}. Such an approach may also help
          explain why mathematics is so "unreasonably
          effective" \cite{Wigner} and why physics is so
          comprehensible \cite{Davies}.

          A basic assumption made here is that quantum
          mechanics or some suitable generalization such as
          quantum field theory is universally applicable.
          One consequence of this assumption is that both
          microscopic and macroscopic systems must be described
          quantum mechanically.  This includes both microscopic
          and macroscopic computers.  The fact that macroscopic
          computers which are in such wide use can be described
          classically to a very good approximation does not invalidate
          a quantum mechanical description by use of pure or mixed states,
          no matter how complex the system may be.

          Because of the universality of quantum mechanics, the interest
          here is in quantum mechanical  models of the
          axiom systems of various types of numbers. The approach taken
          here differs from that usually taken in that emphasis is placed
          on the operations and their properties as described by the axioms
          rather than on the states of a system. For example, the axioms
          for the natural numbers describe three basic operations, the
          successor (corresponding to $+1$), $+$, and $\times$ and their properties.
          A mathematical or physical system is considered to
          represent the natural numbers, or be a model of the axioms if
          the representations of the basic operations satisfy the axioms.

          Here the main emphasis is on mathematical models based on quantum systems.
          The importance of physical models includes the  basic requirement
          that each of the basic operators corresponding to the operations
          described by the axioms must be efficiently
          implementable \cite{BenRNQM}. In brief this requirement
          means that, for each basic operation $O$,
          there must exist a physically implementable quantum
          dynamics that carries out  $O$ on the
          number states. The requirement of efficiency means that the space-time and
          thermodynamic resources needed to implement the operations on states
          representing a number $N$ must be polynomial in $\log_{k}{N}$ where $k\geq 2$.
          In particular the resources required should not be polynomial in
          $N$.

          The requirement of efficient implementation is quite
          restrictive, especially for microscopic quantum systems. (See
          \cite{DiVincenzo,DiV0002077} for a description of some implementation
          conditions for microscopic systems.)
          Quantum systems for which the basic operations are not
          efficiently implementable cannot serve as physical
          models of the axioms being considered. Examples include
          systems with states corresponding to unary
          representations of numbers (for which all arithmetic operations
          are exponentially hard) or in very noisy and chaotic
          environments.

          However it is also the case that there must exist either
          macroscopic or microscopic physical models of the
          axioms of arithmetic. In particular, physical models must
          exist that represent the numbers  $0,1,\cdots ,N$  where $N$
          is large, and are capable of carrying out arithmetic operations
          on these numbers.  If such
          models did not exist, even for moderate values of $N$, it
          would not be possible to carry out any but the most
          elementary calculations or to even develop a physical
          theory of the universe. In a more fundamental sense the
          requirement that there must exist physical models of
          the axioms must in some way place restrictions on the
          basic properties of the physical universe in which we
          live. That is, in some way it must be related to the strong
          anthropic principle  \cite{Anthro,Hogan}.

          The axioms of arithmetic, in common with other
          mathematical axiom systems, make no mention of the requirements of
          efficiency or implementability.  These are extra conditions that
          are essential for the existence of  physical models of
          the axioms \cite{BenRNQM}. They play no role in the existence of mathematical
          models of the axioms.

          It would be desirable to expand these axiom systems to include some
          aspects of efficient implementability.  For
          example, one problem that is not taken into account so far is that efficient
          implementability of the plus $(+)$ and times $(\times)$ operations
          does not follow from efficient implementability of
          the successor operation.  In fact, implementation of the $+$ or $\times$
          operations by iteration of the successor operation is not efficient
          in that addition or multiplication of two numbers, $n,m$
          requires a number of iterations of the successor
          operation that is polynomial in $m$ and $n$ rather than
          in $\log {nm}$. One can require  that each
          of  the three operations are efficiently implementable, but this
          provides no insight or relation between efficient procedures
          for the successor operation and for the plus and times
          operations.

          This problem is taken into account here by defining many
          successor operations rather than just one.
          For natural numbers and integers the successors $S_{j}$
          for $j=1,2,\cdots$ correspond informally to the
          addition of $k^{j-1}$ where $k$ is an arbitrary integer $\geq 2$.
          $S_{1}$ is the usual successor of axiomatic arithmetic
          \cite{Shoenfield,Smullyan}. For rational numbers the indices are
          extended to negative $j$ values.

          The successors are required to satisfy several
          properties.  The most important one is
          \begin{equation}
          S_{j+1} = (S_{j})^{k} \label{logeff}
          \end{equation}
          This equation states that one iteration of $S_{j+1}$ is
          equivalent to $k$ iterations of $S_{j}$. This makes
          quite clear why it is not sufficient to require that
          just $S_{1}$ be efficiently implementable. Instead each
          of the $S_{j}$ must be efficiently implementable.

          The value of these successor operations is that the
          $+$ and $\times$ operations can be defined
          in terms of polynomially many iterations of the successor
          operations. It follows that if the $S_{j}$ are efficiently implementable
          and if any operation consisting of polynomially many iterations of these
          operations is efficiently implementable, then the
          $+$ and $\times$ operations are efficiently implementable.
          This condition is in fact satisfied by all classical
          computers and will have to be  satisfied by any  quantum computer.

          The definition of efficient implementability given
          earlier  applies here.  $S_{j}$ is efficiently
          implementable if there exists a physical procedure for
          implementing $S_{j}$ and the space-time and
          thermodynamic resources needed to carry out the
          procedure are polynomial in $j$. The resources needed should not be
          exponential in $j$. The fact that this condition applies
          to infinitely many $S_{j}$ is not a problem, because
          procedures for arbitrarily large $j$ would be needed
          only asymptotically.  Any operation that is completed in
          a finite time needs only a finite number of the $S_{j}$
          to be efficiently implementable.

          The axiom systems for the different types of numbers can be extended to
          all the successor operations.  For natural numbers, axioms must be added to describe
          other conditions that the successors should satisfy. These include Eq.
          \ref{logeff} and
          the requirement that for each $j$ there are
          many numbers that cannot be obtained by adding
          $k^{j-1}$ to some other number. That is $S_{1}(x)\neq
          \underline{0}$ and $y\leq S_{j}(0) \rightarrow x\neq
          S_{j}(y)$ for all $x,y$.  Here $x,y$ are number
          variables. Other additions give the requirements that for each $j$
          $S_{j}(x)=S_{j}(y)\rightarrow x=y, \;
          x+S_{j}(y)=S_{j}(x+y), x\times S_{j}(y)=x\times y+ x\times S_{j}(0),
          S_{1}(y)=x \rightarrow x\times S_{j}(0)= y\times S_{j}(0) +S_{j}(0),$
          and $x\times S_{1}(0)=x.$  However the discreteness axiom $x\leq
          S_{1}(y)\rightarrow x\leq y \vee x=S_{1}(y)$ holds only
          for $j=1$. The other axioms, \cite{Shoenfield,Smullyan} including Peano's induction
          axiom, are unchanged.

          Here  earlier work on modular arithmetic of $k-ary$ representations of
          length $L$ of the natural numbers \cite{BenRNQM} will be extended  to
          include $k-ary$ representations of  all natural numbers, not just those $<k^{L}$,
          and the integers and rational numbers. The procedure followed here will
          be to give abstract quantum mechanical models of these types of number
          systems. These serve as a convenient common reference
          for discussion of physical models just as abstract
          representations of networks of quantum gates, as in
          \cite{Beckman,Vedral} do for physical quantum gate
          networks.

          Abstract quantum mechanical models for the natural numbers, the
          integers, and the rational numbers are discussed in Sections \ref{TNN}, \ref{TI},
          and \ref{RN}. Definitions  are given for operators for the successor operations
          for each type of number system. Operators for $+$ and
          $\times$ are defined in terms of iterations of
          the successor operators. It is seen that these operators
          satisfy the properties described by the axioms.

          Maps from these abstract models to physical models of
          quantum systems are described in Section \ref{PMAS}.
          Some aspects of the  condition of efficient implementability of the
          basic operations  are briefly  discussed.
          \section{Fermion and Boson Models} \label{FBM}

          Since all numbers of each type are under consideration, the string
          length in the $k-ary$ representation is unbounded.  In
          particular the string length changes as a result of
          various operations on the numbers. This feature is accounted for here by
          constructing quantum mechanical models of these numbers
          and their operations as  multicomponent states  and operators in Fock space.
          The individual string components are represented by bosonic or fermionic annihilation
          and creation operators, $a_{\ell,j}$ and $a^{\dagger}_{\ell,j}$ that
          annihilate or create a system  in the quantum state $|\ell ,j\rangle$.
          For $k-ary$ representations of numbers  the values of
          $j=1,2,\cdots ,$ correspond to the different powers of $k$ and the
          values of $\ell =0,1,\cdots ,k-1$ are multipliers of the corresponding
          powers of $k$.  This is shown in Eq. \ref{numrep} where where $ \underline{s}$
          becomes a function from $1,2,\cdots ,L$
          to $0,1,\cdots ,k-1$ and $2^{j-1}$ is replaced by $k^{j-1}$.

          From a field theory viewpoint, the states
          $a^{\dagger}_{\ell ,j}|0\rangle$ represent single mode
          excitations of the Fermion or Boson field.  These
          correspond to states of single particles or field
          systems. Multiple mode field excitations of the form
          $a^{\dagger}_{ \underline{s}}|0\rangle$  correspond to
          states of $L$ particles with at most one mode, or
          particle, associated with each value of $j$.

          If the values of $j$ denote different space locations, then the
          states $a^{\dagger}_{\underline{s}}
          |0\rangle$ describe excitations with one particle or
          component at locations $1,2,\cdots ,L$ and no particle
          anywhere else.  This model is often used in physical
          multi particle systems where the values of $\ell$ refer to
          such single system properties as different excitation states, spin
          projections, or polarization properties.

          For bosons or fermions the annihilation creation (a-c) operators
          satisfy commutation or anticommutation relations given by
          \begin{eqnarray}
          [a^{\dagger}_{\ell^{\prime} ,j^{\prime}},a_{\ell
          ,j}] = \delta_{\ell^{\prime} j^{\prime},\ell j} \mbox{
          for bosons} \nonumber \\
          \{a^{\dagger}_{\ell^{\prime} ,j^{\prime}},a_{\ell
          ,j}\} = \delta_{\ell^{\prime} j^{\prime},\ell j} \mbox{
          for fermions} \label{acom}
          \end{eqnarray}
          and \begin{eqnarray}
          [a^{\dagger}_{\ell^{\prime} ,j^{\prime}},a^{\dagger}_{\ell
          ,j}]=0= [a_{\ell^{\prime} ,j^{\prime}},a_{\ell,j}]\mbox{
          for bosons} \nonumber \\
           \{a^{\dagger}_{\ell^{\prime},j^{\prime}},a^{\dagger}_{\ell,j}\}  =0=
            \{a_{\ell^{\prime} ,j^{\prime}},a_{\ell,j}\}  \mbox{ for fermions}. \label{acom1}
          \end{eqnarray}
          Here $\{x,y\}=xy+yx$ and $[x,y]=xy-yx$.

          The basis states of interest have the form
          \begin{equation}
          a^{\dagger}_{\underline{s}(L),L}a^{\dagger}_{\underline{s}(L-1),L-1},
          \cdots,a^{\dagger}_{\underline{s}(1),1}|0\rangle
          =a^{\dagger}_{\underline{s}}|0\rangle \label{basst}
          \end{equation}
          where $ \underline{s}$ is any function from $1,2,\cdots
          ,L$ to $0,1,\cdots ,k-1$ with $L$, the length of
          $\underline{s}$, arbitrary. The convention used here
          for all states is that the component a-c
          operators appear in the order of increasing
          values of $j$. Linear superpositions of
          these states have the form $\psi = \sum_{\underline{s}}
          c_{\underline{s}}a^{\dagger}_{\underline{s}}|0\rangle$
          where the sum is over all functions $ \underline{s}$ of
          finite length. The vacuum state $|0\rangle$ is the state
          corresponding to the zero length function.

          The use of fermion and boson systems to carry out quantum
          computation and the representation of fermions as products of
          Pauli operators has been the subject of some discussion in the
          literature \cite{Bravyi,Ortiz,Vlasov}. Here the change of sign
          associated with permutation of fermion a-c operators does not
          cause problems in that the order of creation operators in the states
          $a^{\dagger}_{\underline{s}}|0\rangle$, shown in Eq. \ref{basst}, will be
          maintained in all states considered here.  Also most terms in the
          operators to be defined either have an even number of a-c operators
          that act either at the same place  or
          on the lefthand operators in Eq. \ref{basst}. For those cases where
          the sign change has an effect, operators for fermions will be defined
          differently than for bosons.

          \section{The Natural Numbers} \label{TNN}
          \subsection{The Successor Operators}

          The approach taken here is to consider the states
          $|\underline{s}\rangle = a^{\dagger}_{
          \underline{s}}|0\rangle$ with $a^{\dagger}_{
          \underline{s}}|0\rangle$ given by eq. \ref{basst} as candidate
          natural number states. Operators for the successors, $+$ and
          $\times$ will be defined and seen to have the properties
          specified by the axioms. This shows that, relative to the
          operator definitions, the above states do represent natural numbers.

          It should be noted that states of the form
          $a^{\dagger}_{\underline{s}} |0\rangle$ give a many one
          representation of the numbers in that states with
          arbitrary extensions with $0s$ to the left correspond
          to the same number as those without. ($00134$ is the
          same number as $134$). However the main interest here is
          in states where $s(L)\neq 0$; those with $s({L} =0$ will
          play a role as intermediate states only.

          Let ${\mathcal H}$ be the Hilbert space spanned by all
          states of this form.  Define the operators $P_{occ,j}
          =\sum_{h=0}^{k-1}a^{\dagger}_{h,j}a_{h,j}$ and $P_{>0,j} =
          \sum_{h=1}^{k-1}a^{\dagger}_{h,j}a_{h,j}$.  These
          operators are the number operators for finding a
          particle in any  state $h$ for a fixed value of $j$, and in any
          state $h\neq 0$.  Since ${\mathcal H}$, as a subspace
          of the full Fock space, is defined
          so that at most one component or particle can have property $j$ for either
          fermions or bosons, the eigenvalues of
          these  number operators on ${\mathcal H}$ are just $0,1$.
          Because of this they are shown as projection operators.
          $P_{unocc,j}=1-P_{occ,j}$ is the projection operator for
          finding the site $j$ unoccupied.

          Based on these definitions the successor operators can be defined
          for each value of $j$ as
          \begin{equation}
          V_{j} = N_{j}Z_{j} \label{Vsucc}
          \end{equation}
          where
          \begin{eqnarray}
          N_{j} & = \sum_{h=1}^{k-2}a^{\dagger}_{h+1,j}a_{h,j}+
          a^{\dagger}_{1,j}a_{0,j}P_{occ,j+1} \nonumber \\
          &+  N_{j+1}a^{\dagger}_{0,j}a_{k-1,j}+P_{unocc,j+1}a^{\dagger}_{1,j}P_{unocc,j}
          \label {Nsucc}
          \end{eqnarray}
          for $j\geq 2$ and
          \begin{equation}
          N_{1}= \sum_{h=0}^{k-2}a^{\dagger}_{h+1,1}a_{h,1}+
          N_{2}a^{\dagger}_{0,1}a_{k-1,1}.
          \end{equation}
          $Z_{j}$ is defined as
          \begin{eqnarray}
          Z_{j} & = &P_{occ,j}+P_{unocc,j}P_{>0,j-1} \nonumber \\
          & + & \sum_{\ell=2}^{j-2}a^{\dagger}_{0,j-1},\cdots
          ,a^{\dagger}_{0,\ell +1} P_{unocc,\ell +1}P_{>0,\ell} \nonumber \\
           & + &a^{\dagger}_{0,j-1},\cdots
          ,a^{\dagger}_{0,2}P_{unocc,2}. \label{Zdef}
          \end{eqnarray}
          for $j\geq 4$.  $Z_{1}=1=Z_{2}$ and $Z_{3}$ is obtained from
          Eq. \ref{Zdef} by deleting the sum terms.

          The operator $V_{j}$ is a product of two operators
          $N_{j}$ and $Z_{j}$. The first three terms of $N_{j}$ with
          the first term of $Z_{j}$ act on states $a^{\dagger}_{ \underline{s}}|0\rangle$
          where  site $j$ is occupied.  That is, $a^{\dagger}_{ \underline{s}}$ includes
          a creation operator $a^{\dagger}_{h,j}$ for some value of $h$. The first term of
          $N_{j}$  converts $|h,j\rangle$ to $|h+1,j\rangle$ if $1\leq h\leq k-2$.
          The second term converts $|0,j\rangle$ to $|1,j\rangle$ if site $j+1$  is occupied,
          and the third term converts $|k-1,j\rangle$  to $|0,j\rangle$ with the
          carry one operation shown by the subsequent
          action of $N_{j+1}$.

          The  last term of $N_{j}$ with the remaining three terms of
          $Z_{j}$ act on states where site $j$ is unoccupied. The
          effect of the three terms of $Z_{j}$ acting on $a^{\dagger}_{
          \underline{s}}|0\rangle$, where  the length, $L$, of $
          \underline{s}$ is less than $j$, is to extend $\underline{s}$
          by adding $0s$ so that sites $\leq j-1$ are occupied.
          The action of the last term of $N_{j}$ on $Z_{j}a^{\dagger}_{ \underline{s}}|0\rangle$
          is to create a $1$ at site $j$ just to the left of the leftmost $0$.
          As an example, if $a_{ \underline{s}}|0\rangle = |364\rangle$ and $j=7$, then
          $Z_{7}|364\rangle = |000364\rangle$ and
          $N_{j}Z_{j}|364\rangle = |1000364\rangle$.

          The definition of $Z_{j}$ is explicit and shows the
          operator to be a many system nonlocal operator.  However
          it can also be defined recursively by
          \begin{equation}
          Z_{j} =  P_{occ,j}+ P_{unocc,j}P_{>0,j-1}  + Q_{j-1}
          \end{equation}
          where
          \begin{eqnarray}
          Q_{j-1}& =
          a^{\dagger}_{0,j-1}(P_{unocc,j-1}P_{>0,j-2}+Q_{j-2})
          \nonumber \\
          Q_{2}& =P_{unocc,3}P_{occ,2}+a^{\dagger}_{0,2}P_{unocc,2}. \label{recZ}
          \end{eqnarray}
          This form shows that $Z_{j}$ can also be expressed as a
          product of local operators.  $N_{j}$ is already defined
          in this form although the recursion is in the direction
          of increasing $j$.  The recursion direction does not cause a problem in
          that for any state $a^{\dagger}_{\underline{s}}
          |0\rangle$ with $L>j$, there are at most $L-j+1$
          recursions with $N_{L+1}$ being the last one.

          The recursive forms of both $N_{j}$ and $Z_{j}$ show
          explicitly that these operators, and $V_{j}$, are
          efficiently implementable, relative to that for the
          local a-c operators.  $Z_{j}$ is a sum of products of
          at most $j+1$ local a-c operators with one term in the
          sum active on each number state of the form $a^{\dagger}_{\underline{s}}
          |0\rangle$.  As such it can be implemented in polynomially
          $j$ many steps as its role is to extend a number string by
          adding up to $j-1$ zeros, if needed. The same argument, applied to $N_{j}$
          with at most $L-j+1$ recursions, shows that it can be implemented in
          polynomially $L$ many steps.

          Define the subspace ${\mathcal H}^{arith}$ of ${\mathcal
          H}$ as the Hilbert space spanned by states of the
          form $a^{\dagger}_{\underline{s}} |0\rangle$ where
          $\underline{s}(L)\neq 0$ if $L>1$ and
          $\underline{s}(1)=0,1,\cdots ,k-1$ if $L=1$. Then there
          is a one to one correspondence between these basis states
          and the natural numbers where the state
          $a^{\dagger}_{0,1}|0\rangle$ corresponds to $0$. This
          space is invariant for the $V_{j}$ ($V_{j}{\mathcal
          H}^{arith}\subset {\mathcal H}^{arith}$ even though
          $Z_{j}$ takes states in ${\mathcal H}^{arith}$ outside ${\mathcal
          H}^{arith}$ into ${\mathcal H}\ominus {\mathcal H}^{arith}$).

          If the $V_{j}$ can be shown to satisfy the properties of
          the successor operations given by the (expanded) axioms
          of arithmetic, then they correspond to addition of
          $k^{j-1}$. In this case the adjoint, $V^{\dagger}_{j}$,
          corresponds to subtraction of $k^{j-1}$ on the domain
          of definition. $V^{\dagger}_{j}=Z^{\dagger}_{j}N^{\dagger}_{j}$ where
          \begin{eqnarray}
          N^{\dagger}_{j}& =
          \sum_{h=1}^{k-2}a^{\dagger}_{h,j}a_{h+1,j}+
          P_{occ,j+1}a^{\dagger}_{0,j}a_{1,j} \nonumber \\
          &+a^{\dagger}_{k-1,j}a_{0,j}N^{\dagger}_{j+1}+P_{unocc,j}a_{1,j}P_{unocc,j+1}
          \label {Nsuccdag}
          \end{eqnarray}
          and
          \begin{eqnarray}
          Z^{\dagger}_{j}& = &P_{>0,j}P_{unocc,j}+P_{occ,j}\nonumber \\
          &+ &\sum_{\ell =2}^{j-2} P_{>0,\ell}P_{unocc,\ell
          +1}a_{0,\ell +1},\cdots ,a_{0,j-1} \nonumber \\
          &+ &P_{unocc,2}a_{0,2},\cdots ,a_{0,j-1}.\label{Zdefdag}
          \end{eqnarray}

          Based on the above it can be seen that, relative to the
          space ${\mathcal H}^{arith}$, each
          $V_{j}$ is a unilateral shift \cite{Halmos}.  That is,
          $V^{\dagger}_{j}V_{j} =1$ and
          $V_{j}V^{\dagger}_{j} =P$ where $P$ is a projection
          operator on a subspace of ${\mathcal H}^{arith}$. Also
          for each state $a^{\dagger}_{\underline{s}} |0\rangle
          \equiv | \underline{s}\rangle ,\; \langle
          \underline{s}|V_{j}| \underline{s}\rangle = 0$.

          To see that
          $V^{\dagger}_{j}V_{j}=Z^{\dagger}_{j}N^{\dagger}_{j}N_{j}Z_{j}
          =1$ one notes that $N^{\dagger}_{j}N_{j}
          =\sum_{h=1}^{k-2}P_{h,j} +P_{occ,j+1}P_{0,j}
          +P_{k-1,j}N^{\dagger}_{j+1}N_{j+1}
          +P_{unocc,j}P_{unocc,j+1}$ as only the diagonal terms
          are nonzero. This shows that $\langle
          \underline{s}|N^{\dagger}_{j}N_{j}|\underline{s}\rangle
          = 1$ for all $ \underline{s}$ for which $L =j-1$
          (including $\underline{s}$ for which $
          \underline{s}(L)=0$) and $L\geq j$ and $
          \underline{s}(L)>0$.  Since $Z_{j}$ passes unchanged all
          states $|\underline{s}\rangle$ in ${\mathcal H}^{arith}$
          for which $L\geq j-1$, one has
          $\langle\underline{s}|V^{\dagger}_{j}V_{j}|\underline{s}\rangle
          =1$ for these states.  For states $|\underline{s}\rangle$ in ${\mathcal H}^{arith}$
          for which $L<j-1$, $Z_{j}$ and $N_{j}$ are defined so that
          $N^{\dagger}_{j}N_{j}Z_{j}|\underline{s}\rangle=Z_{j}|\underline{s}\rangle$.
          Thus $\langle\underline{s}|V^{\dagger}_{j}V_{j}|\underline{s}\rangle
          =1$ for these states also which completes the proof.

          Inspection of the terms in $N_{j}Z_{j}$ shows directly
          that $\langle\underline{s}|V_{j}|\underline{s}\rangle
          =0$.  Finally one sees from Eq. \ref{Nsuccdag} that
          $V^{\dagger}_{j}|\underline{s}\rangle =0$ for all $
          \underline{s}$ for which $L\leq j-1$. This shows that
          $V_{j}V^{\dagger}_{j}|\underline{s}\rangle =0$ on these
          states.  For all states $|\underline{s}\rangle$ for
          which $L\geq j$, $V_{j}V^{\dagger}_{j}|
          \underline{s}\rangle =N_{j}N^{\dagger}_{j}|
          \underline{s}\rangle$.  An argument similar to that for
          $V^{\dagger}_{j}V_{j}$ shows that $\langle
          \underline{s}|V_{j}V^{\dagger}_{j}|
          \underline{s}\rangle=1$ on these states. This completes
          the proof that $V_{j}$ is a unilateral shift.

          This result and $\langle\underline{s}|V_{j}|
          \underline{s}\rangle=0$ show that, if the $V_{j}$ and the
          definitions of $=$ and $\times$ operators (Subsections
          \ref{NNplus} and \ref{NNtimes}) satisfy the
          arithmetic axioms, then the candidate number states do
          represent numbers.  This is the reason for referring,
          in the foregoing, to the states $|\underline{s}\rangle
          =a^{\dagger}_{ \underline{s}}|0\rangle$ as number states.

          The most important required property of the $V_{j}$ is
          that given by Eq. \ref{logeff}: or
          \begin{equation}
          (V_{j})^{k}=V_{j+1}. \label{Vlogeff}
          \end{equation}
          To prove this one first notes
          that $(V_{j})^{h}|\underline{s}\rangle=
          (N_{j})^{h}Z_{j}|\underline{s}\rangle$ for $h\geq 1$. To
          save on notation let $V^{h}_{j},\;N^{h}_{j}$ denote
          $(V_{j})^{h},\;(N_{j})^{h}$ respectively.  First note
          that $V^{h}_{j}|\underline{s}\rangle =
          N^{h}_{j}Z_{j}|\underline{s}\rangle$.

          There are several cases to consider.  For
          $|\underline{s}\rangle$ where $L<j-1$
          $$Z_{j}|\underline{s}\rangle=
          |0_{[j-1,L+1]}*\underline{s}\rangle=
          a^{\dagger}_{0,j-1},\cdots
          ,a^{\dagger}_{0,L+1}a^{\dagger}_{
          \underline{s}}|0\rangle .$$  Here $*$ denotes
          concatenation and $0_{[a,b]}$ denotes a string of zeroes
          from a to b. Iteration of $N_{j}$ on
          $Z_{j}|\underline{s}\rangle$ gives
          \begin{eqnarray}
          N_{j}Z_{j}|\underline{s}\rangle &=&|1_{j}*
          0_{[j-1,L+1]}*\underline{s}\rangle \nonumber \\
          N^{k-1}Z_{j}|\underline{s}\rangle &=&|k-1_{j}*
          0_{[j-1,L+1]}*\underline{s}\rangle \nonumber \\
          N^{k-1}_{j}Z_{j}|\underline{s}\rangle &=& N_{j+1}|0_{j}*
          0_{[j-1,L+1]}*\underline{s}\rangle \nonumber \\
          &=&N_{j+1}Z_{j+1}|\underline{s}\rangle =
          V_{j+1}|\underline{s}\rangle. \nonumber
          \end{eqnarray}
          Use of $V^{k}_{j}|\underline{s}\rangle
          =N^{k}_{j}Z_{j}|\underline{s}\rangle$ completes the
          proof for this case.

          The case of $L=j-1$ is similar and is left to the reader.
          For $L\geq j$ write $|\underline{s}\rangle =
          |\underline{s}_{[L,j+1]}*\ell_{j}*\underline{s}_{[j-1,1]}\rangle$
          where $0\leq \ell\leq k-1$. Use of
          $Z_{j}|\underline{s}\rangle =|\underline{s}\rangle$
          gives
          \begin{eqnarray}
          &V^{k-1-\ell}_{j}|\underline{s}\rangle &=
          |\underline{s}_{[L,j+1]}*
          k-1_{j}*\underline{s}_{[j-1,1]}\rangle  \nonumber \\
          &V^{k-\ell}_{j}|\underline{s}\rangle
          &=N_{j+1}|\underline{s}_{[L,j+1]}*
          0_{j}*\underline{s}_{[j-1,1]}\rangle \nonumber \\
          &V^{k}_{j}|\underline{s}\rangle &=
          N^{\ell}_{j}N_{j+1}|\underline{s}_{[L,j+1]}*
          0_{j}*\underline{s}_{[j-1,1]}\rangle \nonumber \\
          & &=N_{j+1}|\underline{s}\rangle \nonumber
          \end{eqnarray}
          This result gives immediately that
          $V^{k}_{j}|\underline{s}\rangle=
          V_{j+1}|\underline{s}\rangle$ which is the desired
          result.  To obtain this use was made of the fact that
          $N_{j}N_{j+1}|\underline{s^{\prime}}\rangle=
          N_{j+1}N_{j}|\underline{s^{\prime}}\rangle$. This holds
          even for fermions because the terms in $N_{j}$ and
          $N_{j+1}$ giving an odd number of creation and
          annihilation operators give zero contribution acting on
          states $|\underline{s^{\prime}}\rangle$ with $L\geq
          j+1$. This completes the proof of Eq. \ref{Vlogeff} as
          all cases have been covered.

          The definitions given so far allow the representation of
          any state $|\underline{s}\rangle $ by
          \begin{equation}
          |\underline{s}\rangle =V^{ \underline{s}(L)}_{L}\cdots
          V^{ \underline{s}(1)}_{1}|\underline{0}\rangle.
          \label{0-to-s}
          \end{equation}
          This relation  is quite useful for proving various properties
          of the arithmetic operators.  It is a special case of
          addition described next. It also serves as a good
          illustration of the fact that, even for fermions, the $V$
          operators with arbitrary subscripts commute.  To see
          this it is sufficient to consider $V_{n}V_{m}$ acting on the
          state $|0\rangle$ as the argument is the same for other states.
          Let $n<m$.  Use of Eqs. \ref{Nsucc} and \ref{Zdef} gives
           $$\begin{array}{l}V_{n}V_{m}|\underline{0}\rangle =V_{n} N_{m}
           a^{\dagger}_{0,m-1}\cdots a^{\dagger}_{0,1}|0\rangle \\
           =V_{n}a^{\dagger}_{1,m}a^{\dagger}_{0,m-1}\cdots
           a^{\dagger}_{0,1}|0\rangle \\
           =a^{\dagger}_{1,n}a_{0,n}P_{occ,n+1}a^{\dagger}_{1,m}a^{\dagger}_{0,m-1}\cdots
           a^{\dagger}_{0,1}|0\rangle .\end{array}$$  Commuting
           the leftmost pair of $a$ operators to
           $a^{\dagger}_{0,n}$ on which they act causes no sign
           change.  This shows that
           $V_{n}V_{m}|\underline{0}\rangle =|\underline{s}\rangle
           =V_{m}V_{n}|\underline{0}\rangle$ where $
           \underline{s}(m)= \underline{s}(n)=1, \;
           \underline{s}(\ell) =0$ for all $1\geq \ell \geq m ,\;
           \ell \neq n,m$.

          \subsection{Addition}\label{NNplus}
          The definition of an addition operator $ \widetilde{+}$ is a
          generalization of Eq. \ref{0-to-s}.  Since $ \widetilde{+}$ is a binary
          operator, it acts on pairs of states
          $|\underline{s}\rangle \otimes
          |\underline{t}\rangle=|\underline{s},\underline{t}\rangle$.\footnote{Various
          methods are available to distinguish the state
          $|\underline{s},\underline{t}\rangle$ from
          $|\underline{s}*\underline{t}\rangle$.  These include use
          of special ending symbols for the end of string states
          or extra degrees of freedom in the parameter set for
          distinguishing the component systems.  However, this
          will not be gone into here.} The action of $
          \widetilde{+}$ can be defined by \cite{BenRNQM}
          \begin{equation}
           \widetilde{+}|\underline{s}\rangle \otimes
          |\underline{t}\rangle =|\underline{s}\rangle \otimes
          |\underline{s}+\underline{t}\rangle \label{defplus}
          \end{equation}
          where
          \begin{equation}
          |\underline{s}+\underline{t}\rangle =V^{
          \underline{s}(L)}_{L}V^{ \underline{s}(L-1)}_{L-1}\cdots
          V^{ \underline{s}(1)}_{1}|\underline{t}\rangle
          \label{Vdefplus}
          \end{equation}

          As defined, $\widetilde{+}$ is an isometry \cite{Halmos}.
          The property $\widetilde{+}^{\dagger}\widetilde{+}=1$
          follows from  the fact that the $V_{j}$ are unilateral
          shifts. That $\widetilde{+}\widetilde{+}^{\dagger}$ is a
          projection operator follows from the fact that the adjoint
          $\widetilde{+}^{\dagger}$, which corresponds
          to subtraction, is defined on states
          $|\underline{s},\underline{t}\rangle$ only if
          $|\underline{s}\rangle \leq |\underline{t}\rangle$.
          This follows from
          $\widetilde{+}^{\dagger}|\underline{s},\underline{t}\rangle
          = |\underline{s},\underline{t}-\underline{s}\rangle$
          where $|\underline{t}-\underline{s}\rangle =
          (V^{\dagger})^{ \underline{s}(1)}_{1}\cdots V^{\dagger})^{
          \underline{s}(L)}_{L}|\underline{t}\rangle$  This state
          is defined if and only if all the iterations of the
          adjoints of the $V_{j}$ are defined on the states on
          which they operate.

          This argument shows that $\widetilde{+}$ is the direct sum of an identity operator
          and a unilateral shift.  It is the identity operator on the subspace
          spanned by all states of the form $|\underline{0}\rangle
          |\underline{t}\rangle$ for any $|\underline{t}\rangle$.
          It is a unilateral shift on the subspace spanned by all
          states $|\underline{s}\rangle
          |\underline{t}\rangle$ where  $|\underline{s}\rangle
          \neq |\underline{0}\rangle$.

          \subsection{Multiplication} \label{NNtimes}

          A definition of multiplication can be given that is
          based on successive iterations of addition and a shift
          operator.  The goal of the shift operator $U$ is to shift a
          state $|\underline{s}\rangle$ to a state
          \begin{equation}
          U|\underline{s}\rangle =|\underline{s}*0\rangle =
          a^{\dagger}_{ \underline{s}(L),L+1}\cdots a^{\dagger}_{
          \underline{s}(1),2}a^{\dagger}_{0,1}|0\rangle
          \label{defshift} \end{equation}
          This corresponds informally to multiplying $
          \underline{s}$ by $k$.

          This operator consists of two parts: shifting a state and
          insertion of a $0$ at site $1$. Because the insertion
          involves a single creation operator, the operator must
          be defined differently for fermions than for bosons.
          This is the first case where this distinction matters;
          as was noted the definitions of both the successor and addition
          operators  were the same for both boson and fermion
          states in ${\mathcal H}^{arith}$.  The definition of the shift
          operator $U^{i}$ for both fermions, $i=f$, and bosons, $i=b$, can be given by
          \begin{equation}
          U^{i}=\sum_{j=1}^{\infty}U^{i}_{j}P_{unocc,j+1} \label{Usumi}
          \end{equation}
          where
          \begin{eqnarray}
          U^{i}_{j} &=&
          sg(i)U^{f}_{j-1}\sum_{h=0}^{k-1}a^{\dagger}_{h,j+1}a_{h,j}
          \nonumber \\
          U_{1} &=&
          \sum_{h=0}^{k-1}a^{\dagger}_{h,2}a^{\dagger}_{0,1}a_{h,1}.
          \label{defUi} \end{eqnarray}
          Here $sg(i) = -1$ if $i=f$ and $sg(i) = +1$ if $i=b$.
          $P_{unocc,j}=1-P_{occ,j}=
          1-\sum_{h=0}^{k-1}a^{\dagger}_{h,j}a_{h,j}$ is the
          projection operator for finding no component system at
          site $j$. $P_{occ,j}$ is the site $j$ number operator.
          It is a projection operator on ${\mathcal H}^{arith}$ as
          the only possible eigenvalues are $0$ and $1$ for both
          boson and fermion states. (Recall the definition of ${\mathcal
          H}^{arith}$ as the space spanned by all states of the
          form $|\underline{s}\rangle$ with $ \underline{s}(L)>0$ if $L>1$.)

          The presence of $P_{unocc,j}$ in the definition gives
          the result that $U^{i}|\underline{s}\rangle
          =U^{i}_{L}|\underline{s}\rangle$ where $i$ denotes either
          $f$ or $b$. For fermions the presence of the minus sign
          in Eq. \ref{defUi} means that when $U_{1}$ becomes
          active  on the state $a^{\dagger}_{ \underline{s}(L),L+1}
          \cdots a^{\dagger}_{ \underline{s}(2),3}a^{\dagger}_{ \underline{s}(1),1}
          |0\rangle$ it is multiplied by a factor of $(-1)^{L-1}$.
          There are $3(L-1)$ commutations of the three operators
          in $U_{1}$ to their action on $\cdots a^{\dagger}_{ \underline{s}(1),1}
          |0\rangle$ which gives a total factor of $(-1)^{4L-4}$
          which is positive for any $L$. This shows that Eq.
          \ref{defshift} is satisfied for any
          $|\underline{s}\rangle$ for either bosons or fermions.

          As defined, $U^{i}$ is an isometry.  That is, in
          ${\mathcal H}^{arith}$
          $(U^{i})^{\dagger}U^{i}=1$ but
          $U^{i}(U^{i})^{\dagger}=P_{0,1}P_{occ,2}$.  Here $P_{0,1}$ is the
          projection operator on all states
          $|\underline{s}\rangle$ such that $ \underline{s}(1)=0$.
           This follows from the fact that
           $(U^{i})^{\dagger}|\underline{s}\rangle = \cdots
           (U_{1})^{dagger}|\underline{s}\rangle = \cdots
           \sum_{h=0}^{j-1}a^{\dagger}_{h,1}a_{0,1}a_{h,2}
           \underline{s}\rangle = 0$ unless $ \underline{s}(1)=0$.

          The multiplication operator $ \widetilde{\times}$ is
          defined on triples of states by
          \begin{equation}
          \widetilde{\times}|\underline{s}\rangle\otimes
          |\underline{t}\rangle\otimes|\underline{x}\rangle =|\underline{s}\rangle\otimes
          |\underline{t}\rangle\otimes|\underline{x}+\underline{s}\times
          \underline{t}\rangle.
          \label{deftimes}
          \end{equation}
          Informally the operation multiplies $ \underline{s}$ and
          $ \underline{t}$ and adds the result to $\underline{x}$.
          Pure multiplication occurs when $|\underline{x}\rangle =
          a^{\dagger}_{0,1}|0\rangle$.

          The operator $ \widetilde{\times}$ is expressed in terms
          of $U^{i}$ and $ \widetilde{+}$ by
          \begin{eqnarray}
          \widetilde{\times}|\underline{s},
          \underline{t},\underline{x}\rangle
          &=&((U^{i}_{2})^{\dagger})^{L-1}(\widetilde{+}_{2,3})^{
\underline{s}(L)}U^{i}_{2}
          (\widetilde{+}_{2,3})^{ \underline{s}(L-1)}
          U^{i}_{2} \nonumber \\ & \mbox{} & \cdots U^{i}_{2}
          (\widetilde{+}_{2,3})^{ \underline{s}(1)}
          |\underline{s},\underline{t},\underline{x}\rangle
          \label{deftimes1}
          \end{eqnarray}
          In this equation $i=f,b$ and the subscripts $2,3$ on
          $\widetilde{+}_{2,3}$ and $2$ on $U^{i}_{2}$ show that
$\widetilde{+}_{2,3}=1\otimes
          \widetilde{+}$ and $U^{i}_{2} = 1\otimes U^{i}\otimes
          1$. Also $|\underline{s}\rangle\otimes |\underline{t}\rangle\otimes
          |\underline{x}\rangle = |\underline{s},\underline{t},\underline{x}\rangle$.
          The number of iterations of each plus operation is
          determined by the elements of $|\underline{s}\rangle$.
          Informally the action of $ \widetilde{\times}$ can be
          characterized by  $ \underline{s}(1)$
          iterations of adding $ \underline{t}$ to $
          \underline{x}$, then the addition of $ \underline{s}(2)$
          iterations of adding $ k\underline{t}$ to the result, then $\cdots$,
          then the addition of $ \underline{s}(L)$
          iterations of adding $ k^{L-1}\underline{t}$ to the
          result.  The factor $((U^{i}_{2})^{\dagger})^{L-1}$ restores the
          state $|\underline{k^{L-1}t}\rangle$ to $|\underline{t}\rangle$.

          As is the case for $ \widetilde{+}$, $
          \widetilde{\times}$ is an isometry that is the direct sum of an
          identity and a unilateral shift. It is the identity operator on
          the subspace spanned by all states $|\underline{s}\rangle\otimes
          |\underline{t}\rangle\otimes|\underline{x}\rangle$ where
          either $|\underline{s}\rangle$ or $|\underline{t}\rangle$
          equal $|\underline{0}\rangle$. It is a unilateral shift
          on the subspace spanned by all states $|\underline{s},\underline{t},
          \underline{x}\rangle$ where
          $|\underline{s}\rangle\neq |\underline{0}\rangle \neq
          |\underline{t}\rangle$. On this subspace $\langle\underline{s},
          \underline{t},\underline{x}|\widetilde{\times}
          |\underline{s},\underline{t},\underline{x}\rangle=0$.

          \subsection{The Arithmetic Axioms}

          It is easy to show, based on the properties given
          above, that the operators $V_{j}$,  $\widetilde{+}$, and $ \widetilde{\times}$
          satisfy the arithmetic axioms for the successors and
          plus. That $|\underline{0}\rangle$ is the additive
          identity follows from Eqs. \ref{defplus} and
          \ref{Vdefplus} and is expressed in Eq. \ref{0-to-s}.
          Eq. \ref{Vlogeff} has already been proved.  Also $|\underline{1}\rangle$
          is the multiplicative identity, as can be seen from $ \widetilde{\times}
          |\underline{1}\rangle\otimes|\underline{t}\rangle\otimes
          |\underline{0}\rangle = \widetilde{+}
          |\underline{1}\rangle\otimes|\underline{t}\rangle\otimes
          |\underline{0}\rangle =
          |\underline{1}\rangle\otimes|\underline{t}\rangle\otimes
          |\underline{t}\rangle$. The
          commutativity of the $V_{j}$ and $\widetilde{+}$, or
          $\widetilde{+} |\underline{s}\rangle\otimes
          V_{j}|\underline{t}\rangle =(\widetilde{1}\otimes V_{j})
          \widetilde{+}|\underline{s}\rangle\otimes
          |\underline{t}\rangle$ follows from the definitions of
          the operators involved.

          Again there are no problems even
          for fermions because any terms with an odd number of
          annihilation or creation operators which give a nonzero
          contribution undergo an even number of permutations to
          arrive at the point of action (i.e. where the delta
          functions of Eqs. \ref{acom} and \ref{acom1} apply).

          Proofs of the other two axioms, $ \widetilde{\times}
          (V_{j}\otimes \widetilde{1}\otimes\widetilde{1})|\underline{s}\rangle\otimes
          |\underline{t}\rangle\otimes|\underline{x}\rangle =|\underline{s}\rangle\otimes
          |\underline{t}\rangle\otimes|\underline{y}\rangle$ where
          $|\underline{y}\rangle
          =|\underline{x}+\underline{s}\times
          \underline{t}+\underline{k^{j-1}t}\rangle$, and
          distributivity of multiplication over addition, are
          discussed in the Appendix.

          \section{The Integers} \label{TI}

          As is well known the integers correspond to positive and
          negative natural numbers.  A suitable set of axioms can
          be obtained by replacing the arithmetic axiom $"0\neq
          S(x)"$ ($0$ is not a successor of any element) by
          $"\forall{x}\exists{y}(x=S(y))"$ (every element is a
          successor). Also an axiom stating the existence of an
          inverse to addition is needed. As was done for the
          natural numbers extension of these axioms to include
          all the successors $S_{1},S_{2},\cdots$ is needed.
          Integers also satisfy the axioms for a commutative ring with identity\cite{Adamson}.

          Here integers will be represented by states of the form
          \begin{eqnarray}
          |+ \underline{s}\rangle &=
          &a^{\dagger}_{+,L+1}a^{\dagger}_{ \underline{s}}|0\rangle
          \nonumber \\
          |- \underline{s}\rangle &=
          &a^{\dagger}_{-,L+1}a^{\dagger}_{ \underline{s}}|0\rangle
          \label{defint}
          \end{eqnarray}
          where $a^{\dagger}_{ \underline{s}}|0\rangle =
          a^{\dagger}_{ \underline{s}(L),L}\cdots a^{\dagger}_{
          \underline{s}(1),1}|0\rangle$ is the same definition as
          was used for the natural numbers. By convention the
          integer $0$ will be represented by the positive version
          only, or $a^{\dagger}_{+,2}a^{\dagger}_{0,1}|0\rangle =
          |+\underline{0}\rangle$.  The Hilbert space of interest,
          ${\mathcal H}^{I}$ is spanned by all states
          $|\pm\underline{s}\rangle$ where $ \underline{s}(L)\neq
          0$ and the state $|+\underline{0}\rangle$. This is a
          subspace of a space that includes states of the form
          $|\pm\underline{s}\rangle$ where $ \underline{s}(L)=0$
          is possible.

          As was the case for the natural numbers this definition
          is valid for either bosons or fermions.  In the latter
          case the order of creation operators appearing in Eq.
          \ref{defint}, that mirrors the  ordering of the site
          labels for the component systems (with the sign
          component at the end), is taken to be fixed.

          \subsection{The Successor Operators} \label{ISO}

          As was the case for the natural numbers, successor
          operators $I_{j}$ are defined, one for each
          $j=1,2,\cdots$, that are to correspond to addition of
          $k^{j-1}$.  $I_{j}$ consists of three components, one
          for the nonnegative integers and two for the negative
          integers separated on the basis of whether a sign change
          does or does not occur. To this end define the
          projection operators
          \begin{eqnarray}
          P_{+}&=&\sum_{j=2}^{\infty}a^{\dagger}_{_+,j}a_{+,j}
          \nonumber \\
          P_{-,\geq j}
          &=&\sum_{h=j+1}^{\infty}a^{\dagger}_{_-,h}a_{-,h},
          \nonumber \\
          P_{-,< j}
          &=&\sum_{h=2}^{j}a^{\dagger}_{_-,h}a_{-,h}. \label{proj+-}
          \end{eqnarray}
          These are defined as number operators.  On  ${\mathcal
          H}^{I}$ they are projection operators as $0,1$ are
          the only possible eigenvalues. Note that the subscripts $\geq ,\;<$
          refer to the sites of the single digit number operators and
          do not include the sites of the signs. A sign change
          operator $W$ is defined as
          \begin{equation}
          W=\sum_{j=1}^{\infty}(a^{\dagger}_{+,j}a_{-,j} +a^{\dagger}_{-,j}a_{+,j}).
          \label{sgnch} \end{equation}
          $W$ is unitary and $W^{2} =1$.

          The successor operation $I_{j}$ on the Hilbert space ${\mathcal
          H}^{I}$ can be separated into two operators as
          \begin{equation}
          I_{j} = I^{+}_{j}+I^{-}_{j} = I^{+}_{j}+I^{-}_{\geq j}+ I^{-}_{<j}  \label{Idef}
          \end{equation}
          where $I^{+}_{j}$ and $I^{-}_{j}$ are defined on ${\mathcal
          H}^{I+}$ and ${\mathcal H}^{I-}$, the spaces of
          nonnegative and negative integer states respectively.
          $I^{-}_{<j}=I^{-}_{j}P_{-,<j}$ is defined on the subspace $P_{-,<j}{\mathcal
          H}^{I-}$ and $I^{-}_{\geq j}=I^{-}_{j} P_{-,\geq j}$ is defined
          on $P_{-,\geq j}{\mathcal H}^{I-}$. The action of $I^{-}_{<j}$ takes negative
          number states into positive number states. The sign is
          unchanged by the action of $I^{-}_{\geq j}$. Informally
          these two correspond to the addition of $k^{j-1}$ to
          negative numbers whose absolute value is $<k^{j-1}$ and
          $\geq k^{j-1}$ respectively.

          The definitions of the $I^{+}_{j}$ are quite similar to those for the
          natural numbers.  Corresponding to Eqs.
          \ref{Vsucc},\ref{Nsucc}, and \ref{Zdef} one has
          \begin{equation}
          I^{+}_{j} = K^{+}_{j}Z^{+}_{j} \label{Isucc}
          \end{equation}
          where
          \begin{eqnarray}
          K^{+}_{j} & = \sum_{h=1}^{k-2}a^{\dagger}_{h+1,j}a_{h,j}+
          a^{\dagger}_{1,j}a_{0,j}P_{nocc,j+1} \nonumber \\
          &+
          K^{+}_{j+1}a^{\dagger}_{0,j}a_{k-1,j}+a^{\dagger}_{+,j+1}
          a^{\dagger}_{1,j}P_{unocc,j} \label{Ksucc}
          \end{eqnarray}
           for $j\geq 2$ and
           \begin{equation}
           K^{+}_{1} = \sum_{h=0}^{k-2}a^{\dagger}_{h+1,1}a_{h,1}+
          K^{+}_{2}a^{\dagger}_{0,1}a_{k-1,1}. \label{K1succ}
          \end{equation}
          For $j\geq 4$ $Z_{j}$ is defined by
          \begin{eqnarray}
          Z^{+}_{j} & = &P_{unocc,j}a_{+,j}P_{>0,j-1}+P_{nocc,j}P_{+} \nonumber \\
          & + & \sum_{\ell=2}^{j-2}a^{\dagger}_{0,j-1},\cdots
          ,a^{\dagger}_{0,\ell +1} a_{+,\ell +1}P_{>0,\ell} \nonumber \\
           & + &a^{\dagger}_{0,j-1},\cdots
          ,a^{\dagger}_{0,2}a_{+,2}. \label{Z+def}
          \end{eqnarray}
          $Z_{1}=P_{+},\; Z_{2}=P_{nocc,2}P_{+}+a_{+,2}$, and $Z_{3}$ is given by Eq.
          \ref{Z+def} by deleting the sum terms.  $P_{nocc,j+1} =
          \sum_{h=0}^{k-1}a^{\dagger}_{h,j+1}a_{h,j+1}$ is the
          projection operator for site $j+1$ occupied by a system
          in a single digit number state (not in a sign state).
          $P_{unocc,j}$ is the projection operator for site $j$ to
          be unoccupied by a system in any state.

          The operators $K^{+}_{j}$ and $Z^{+}_{j}$ serve the same
          function on the the Hilbert space ${\mathcal H}^{I+}$ of nonnegative integer
          states as do the operators $N_{j}$ and $Z_{j}$ on
          ${\mathcal H}^{arith}$ for the natural number states. As
          was done for $Z_{j}$ in Eq. \ref{recZ}, $Z^{+}_{j}$ can
          also be expressed recursively.

          One can also prove that $I^{+}_{j}$ is a unilateral shift on
          ${\mathcal H}^{I+}$ and that
          \begin{equation}
          I^{+}_{j+1}= (I^{+}_{j})^{k}. \label{logeffI+}
          \end{equation}
          The proofs will not be given here as they are quite
          similar to those for $V_{j}$ given earlier. It is also clear
          that, corresponding to Eq. \ref{0-to-s} one has
          \begin{eqnarray}
          |+\underline{s}\rangle &= (I^{+}_{L})^{
          \underline{s}(L)}(I^{+}_{L-1})^{
          \underline{s}(L-1)},\cdots ,(I^{+}_{1})^{
          \underline{s}(1)}|+0\rangle \nonumber \\ \mbox{} & = I^{+}_{
          \underline{s}}|+0\rangle. \label{0-to-+s}
          \end{eqnarray}

          The adjoint $(I^{+}_{j})^{\dagger}$ is given by $(I^{+}_{j})^{\dagger}
          =(Z^{+}_{j})^{\dagger}(K^{+}_{j})^{\dagger}$ where
           \begin{eqnarray}
          (K^{+}_{j})^{\dagger} & = \sum_{h=1}^{k-2}a^{\dagger}_{h,j}a_{h+1,j}+
          P_{nocc,j+1}a^{\dagger}_{0,j}a_{1,j} \nonumber \\
          &+
          a^{\dagger}_{k-1,j}a_{0,j}(K^{+}_{j+1})^{\dagger}+P_{unocc,j}a_{1,j}
          a_{+,j+1} \label {Idagsucc}
          \end{eqnarray}
           for $j\geq 2$ and
           \begin{equation}
           (K^{+}_{1})^{\dagger} = \sum_{h=0}^{k-2}a^{\dagger}_{h,1}a_{h+1,1}+
          a^{\dagger}_{k-1,1}a_{0,1}(K^{+}_{2})^{\dagger}.
          \end{equation}
          Also
          \begin{eqnarray}
          (Z^{+}_{j})^{\dagger} & = &P_{>0,j-1}a^{\dagger}_{+,j}P_{unocc,j}
          +P_{nocc,j}P_{+} \nonumber \\
          & + & \sum_{\ell=2}^{j-2}P_{>0,\ell}a^{\dagger}_{+,\ell +1}
          a_{0,\ell +1},\cdots ,a_{0,j-1}, \nonumber \\
           & + &a^{\dagger}_{+,2},a_{0,2},\cdots, a_{0,j-1}. \label{+0dagext}
          \end{eqnarray}
          $(Z^{+}_{1})^{\dagger}=P_{+},\;
(Z^{+}_{2})^{+}=P_{nocc,2}P_{+}+a^{\dagger}_{+,2}$,
          and $Z_{3}$ is given by Eq. \ref{+0dagext} by deleting the sum terms.

          As was the case for $V_{j}$ $(I^{+}_{j})^{\dagger}$
          corresponds to subtraction of $k^{j-1}$ on its domain of
          definition.  One way to see this is to note that Eq. \ref{logeffI+}
          can be used to expand $I^{+}_{j+n}$ as a product of
          powers as $I{+}_{j+n}=(I^{+}_{j})^{k}(I^{+}_{j+1})^{k-1}
          \cdots (I^{+}_{j+n-1})^{k-1}$.  Use of
          $(I^{+}_{j})^{\dagger}I^{+}_{j} = 1$ gives the useful
          result
          \begin{equation}
          (I^{+}_{j})^{\dagger}I^{+}_{j+n} =
          (I^{+}_{j})^{k-1}\cdots (I^{+}_{j+n-1})^{k-1}.
          \label{999+}
          \end{equation}
          This is the operator form of the numerical fact that, for
          example, 10000-10=9990 in decimal notation (k=10).

          The adjoint of $I^{+}_{j}$ can be used to define $I^{-}_{\geq j}$ by
          \begin{equation}
          I^{-}_{\geq j}=(P_{-,>0}W+P_{+,0})(I^{+}_{j})^{\dagger}
          WP_{-,\geq j} \label{I-geqj}
          \end{equation}
          with $W$ given by Eq. \ref{sgnch}. Here $P_{+,0}=|+0\rangle
          \langle 0+|$ and $P_{+,>0}$ is the projection operator
          all positive integer states.  For the quantum states of
          interest here, this equation expresses the simple fact that
          if $-m$ is a negative integer with $|m|\geq k^{j-1}$,
          then $-m+k^{j-1}=-(m-k^{j-1})$ and $-k^{j-1}+k^{j-1}= +0$.

          The operator $I^{-}_{<j}$ can be defined by
          \begin{equation}
          I^{-}_{<j}=\sum_{
          \underline{s}}^{L<j}(I^{+}_{\underline{s}})^{\dagger}I^{+}_{j}
          (I^{+}_{\underline{s}})^{\dagger}WP_{-,\underline{s}}.
          \label{I-<j}
          \end{equation}
          In this equation $P_{-,\underline{s}} =
          |-\underline{s}\rangle\langle \underline{s}-|$ is the
          projection operator on the state
          $|-\underline{s}\rangle$,
          $(I^{+}_{\underline{s}})^{\dagger}W_{-+}P_{-\underline{s}}=
          |+0\rangle\langle \underline{s}-|$ converts
          $|-\underline{s}\rangle$ to  $|+0\rangle$, and
          $(I^{+}_{\underline{s}})^{\dagger}I^{+}_{j}|+0\rangle$
          gives the state corresponding to addition of $k^{j-1}$
          to $0$ and subtracting $ \underline{s}$.  The sum is
          over all $ \underline{s}$ whose length $L$ (excluding the
          sign) is less than $j$ and for which $ \underline{s}(L)>0$.

          It is straightforward to see that $I_{j}$, defined by  Eq.
          \ref{Idef} is a bilateral shift.  $I^{\dagger}_{j}I_{j}=1$
          follows from the result that
          \begin{eqnarray}I^{\dagger}_{j}I_{j} &=&
          (I^{+}_{j})^{\dagger}I^{+}_{j}+(I^{-}_{\geq j})^{\dagger}I^{-}_{\geq j}
          + (I^{-}_{< j})^{\dagger}I^{-}_{< j} \nonumber \\ \mbox{} &=& P_{+}+P_{-,\geq
          j}+P_{-,<j}=1.\nonumber \end{eqnarray}
          Here Eqs. \ref{Isucc}, \ref{I-geqj}, and \ref{I-<j} have been
          used. The sum of the projection operators gives the
          identity on ${\mathcal H}^{I}$. In a similar fashion one
          can show that $I_{j}I^{\dagger}_{j}=1$.  $\langle
          \pm\underline{s}|I_{j}|\pm\underline{s}\rangle =0$ for
          all states $|+\underline{s}\rangle$ and
          $|-\underline{s}\rangle$ follows directly from the definition of
          $I_{j}$.

          $I_{j}$ also satisfies
          \begin{equation}
          I_{j+1}=(I_{j})^{k}. \label{logeffI}
          \end{equation}  The proof of this is given in the
          Appendix. One has for $I_{j}$ a generalization of Eq. \ref{999+}:
            \begin{equation}
            (I_{j})^{\dagger}I_{j+n} = (I_{j})^{k-1}\cdots
            (I_{j+n-1})^{k-1}. \label{999I}
            \end{equation}
            Corresponding to Eq. \ref{0-to-+s} one has
            \begin{eqnarray}
            |\pm\underline{s}\rangle &=&(I_{L})^{\pm
          \underline{s}(L)}(I_{L-1})^{\pm
          \underline{s}(L-1)},\cdots ,(I_{1})^{\pm
          \underline{s}(1)}|+0\rangle \nonumber \\  & = & I_{\pm
          \underline{s}}|+0\rangle. \label{0-to+-s}
          \end{eqnarray}
          Here either the plus sign or the minus sign holds
          throughout.  Also $(I_{h})^{-1}=I_{h}^{\dagger}$.

            \subsection{Integer Addition and
            Multiplication} \label{IAM}
            The definition of addition of integers is similar to
            that given for natural numbers. In essence it is a
            generalization of Eq. \ref{0-to+-s}. One
            has\footnotemark[1]
            \begin{equation}
            \widetilde{+}_{I}|\pm\underline{s}\rangle\otimes
            |\pm\underline{t}\rangle = |\pm\underline{s}\rangle\otimes
            |\pm\underline{t}+\pm\underline{s}\rangle
            \label{defplusI}
            \end{equation}
            where
            \begin{eqnarray}
            |\pm\underline{t}+\pm\underline{s}\rangle &=&(I_{L})^{\pm
          \underline{s}(L)},\cdots ,(I_{1})^{\pm
          \underline{s}(1)}|\pm\underline{t}\rangle  \nonumber \\  &=&  I_{\pm
          \underline{s}}|\pm\underline{t}\rangle.
          \label{defplusIpm}
          \end{eqnarray}
          One sees from the definitions, including Eq.
          \ref{0-to+-s}, that $ \widetilde{+}_{I}$ has the correct
          sign properties.  Acting on the state
          $|-\underline{s}\rangle|\pm\underline{t}\rangle$, the
          negative exponents in the above show that the action of $
          \widetilde{+}_{I}$ corresponds to a subtraction of
          $|+\underline{s}\rangle$ from
          $|\pm\underline{t}\rangle$.

          $ \widetilde{+}_{I}$ has been defined so that it is
          unitary: $ \widetilde{+}_{I}^{\dagger} \widetilde{+}_{I}=1=
          \widetilde{+}_{I}\widetilde{+}_{I}^{\dagger}$.  Here
          $\widetilde{+}_{I}^{\dagger}$ corresponds to the
          subtraction operation on integers. Note that
          $\widetilde{+}_{I}^{\dagger}|-\underline{s}\rangle|\pm\underline{t}\rangle
          =|\pm\underline{t}-(-\underline{s}\rangle$ where
          \begin{eqnarray}
            |\pm\underline{t}-(-\underline{s})\rangle
            &=(I^{\dagger}_{1})^{-
          \underline{s}(1)},\cdots ,(I^{\dagger}_{L})^{-
          \underline{s}(L)}|\pm\underline{t}\rangle  \nonumber \\  &=
          I^{\dagger}_{-
          \underline{s}}|\pm\underline{t}\rangle.
          \label{defplusIdagpm}
          \end{eqnarray}
          Since $(I^{\dagger}_{j})^{-\ell} = (I_{j})^{\ell}$ and the
          various $I_{j}$ factors can be applied in any order, this expresses the
          fact that subtraction of $|-\underline{s}\rangle$ corresponds to addition
          of $|+\underline{s}\rangle$.

          The definition of multiplication for the natural
          numbers, Eqs. \ref{deftimes} and \ref{deftimes1}, can be taken over to
          describe integer multiplication:
          \begin{equation}
          \widetilde{\times}|\pm\underline{s},\pm
          \underline{t},\pm\underline{x}\rangle =|\pm\underline{s},
          \pm\underline{t},\pm\underline{x}+(\pm\underline{s}\times\pm
          \underline{t})\rangle.
          \label{defItimes}
          \end{equation}  More explicitly one has
          \begin{eqnarray}
          \widetilde{\times}|\pm \underline{s},\pm \underline{t},\pm \underline{x}\rangle
          =((U^{I,i}_{2})^{\dagger})^{L-1}(\widetilde{+}_{2,3})^{\pm
\underline{s}(L)}U^{I,i}_{2}
          \nonumber \\ \times (\widetilde{+}_{2,3})^{\pm \underline{s}(L-1)}
          U^{I,i}_{2} \cdots U^{I,i}_{2}
          (\widetilde{+}_{2,3})^{\pm \underline{s}(1)}
          |\pm\underline{s},\pm\underline{t},\pm\underline{x}\rangle.
          \label{defItimes1}
          \end{eqnarray}
          Here the main change in the definition is that $
          \widetilde{+}_{2,3}$ corresponds to integer addition
          given by Eq. \ref{defplusI}. Also $U^{I,i}_{2}$ is  defined
          slightly differently than for the natural numbers.  Eq.
          \ref{Usumi} is replaced by \begin{equation}
          U^{I,i} =\sum_{j=1}^{\infty}
          U^{i}_{j}a^{\dagger}_{\pm,j+2}a_{\pm,j+1} \label{UIsum}
          \end{equation} with the definition of $U^{i}_{j}$
          unchanged and given by Eq. \ref{defUi}  for $i=f,b$.
          The change shown above shifts the sign qubits before the
          numeral qubits are shifted. As was the case for the
          natural numbers, $U^{I,i}$ is an isometry.

          It is straightforward to show that the operator $
          \widetilde{\times}$ is unitary.  This follows from the
          results that $ \widetilde{+}_{2,3}$ is unitary and that
          $U^{I,i}_{2}(U^{I,i}_{2})^{\dagger}$ always acts on
          states on which this operator is the identity. It does
          not follow from this that division is defined as it is
          not the inverse of $\widetilde{\times}$. A correct
          definition of  a division operator $ \widetilde{\div}$
          would have to satisfy the requirement that for each pair
          of states $|\pm\underline{s}\rangle ,
          |\pm\underline{x}\rangle$ there is a unique state
          $|\pm\underline{t}\rangle
          =|\pm\underline{x}\div\pm\underline{s}\rangle$ such that
          \begin{equation}
          \widetilde{\div}|\pm\underline{s},\pm
          \underline{t},\pm\underline{x}\rangle =|\pm\underline{s},
          \pm\underline{t},0\rangle.
          \label{defIdiv}
          \end{equation}

          \subsection{The Integer Axioms}\label{TIA}
          The proofs that the  operators $I_{j},\; \widetilde{+},\;
          \widetilde{\times}$ satisfy the axioms for integers is
          quite similar to those for the natural numbers and will
          not be repeated here.  The proof that every element is a
          $j-$ successor, or for each integer state
          $|\underline{x}\rangle$ there is an integer state
          $|\underline{y}\rangle$ such that
          $I_{j}|\underline{y}\rangle =|\underline{x}\rangle$
          follows from the fact that $I_{j}$ is a bilateral shift
          where $|\underline{y}\rangle = I^{\dagger}_{j}|\underline{x}\rangle$.
          Here $|\underline{x}\rangle, |\underline{y}\rangle$
          denote states of the form $|\pm \underline{s}\rangle ,\;
          |\pm \underline{t}\rangle$ with the sign included. The
          existence of an additive inverse follows immediately
          from the unitarity of $ \widetilde{+}$.

          The proof that the various ring axioms are satisfied is
          straightforward. It is of interest to note that proof of
          the commutativity and associativity of addition and
          multiplication for the operators implies the
          corresponding properties for the numbers appearing in
          the exponents. This property, which was noted before
          \cite{BenRNQM} is a consequence of the string character
          or tensor product representation of the integers.

          For example to prove that
          $|\pm\underline{s}+\pm\underline{t}\rangle
          =|\pm\underline{t}+\pm\underline{s}\rangle$ one  uses
          Eqs. \ref{defplusIpm}, \ref{0-to+-s}, and the commutativity
          of the $I$ operators to obtain  $$ \begin{array}{l}
          |\pm\underline{s}+\pm\underline{t}\rangle =
          |\pm\underline{s}\rangle\otimes
          (I_{L_{s}})^{\pm \underline{s}(L_{s})}\cdots
          (I_{1})^{\pm\underline{s}(1)}|\underline{t}\rangle \\ \\=
          |\pm \underline{s}\rangle \otimes (I_{L_{s}})^{\pm
          \underline{s}(L_{s})}\cdots
          (I_{L_{t}+1})^{\pm \underline{s}(L_{t}+1)} \\ \\
          (I_{L_{t}})^{\pm\underline{s}(L_{t})+\pm\underline{t}(L_{t})}
          \cdots(I_{1})^{\pm\underline{s}(1)+\pm\underline{t}(1)}|\underline{+0}\rangle.
          \end{array}$$  Here $L_{t}\leq L_{s}$ has been used.
          One now uses the commutativity of the
          numbers in the exponents to set $(I_{j})^{
          \underline{s}(j)+\underline{t}(j)} = (I_{j})^{
          \underline{t}(j)+\underline{s}(j)}$ for $1\geq j\geq
          L_{t}$ and write $$\begin{array}{l}(I_{L_{s}})^{\pm \underline{s}(L_{s})}\cdots
          (I_{1})^{\pm\underline{s}(1)}|\underline{t}\rangle =
          (I_{L_{t}})^{\pm \underline{t}(L_{t})}\cdots
          (I_{1})^{\pm\underline{t}(1)}|\underline{s}\rangle \\
          =|\pm\underline{t}+\pm\underline{s}\rangle \end{array}$$ which proves
          commutativity.

          A similar situation exists for associativity.  The proof
          of $|\pm\underline{s}+( \pm\underline{t}+\pm
          \underline{w})\rangle = |(\pm
          \underline{s}+\pm\underline{t})+\pm\underline{w}\rangle$ uses
          the equality $(I_{j})^{\pm\underline{s}(j)}+
          \{(I_{j})^{\pm\underline{t}(j)}+(I_{j})^{\pm\underline{w}(j)}\}
          =\{(I_{j})^{\pm\underline{s}(j)}+
          (I_{j})^{\pm\underline{t}(j)}\}+(I_{j}^{\pm\underline{w}(j)}$.
          Proofs for commutativity and associativity for
          multiplication are more involved because of the relative
          complexity of the definition of the $
          \widetilde{\times}$ operator.  However the same ideas
          apply. These will be discussed more later on.

          \section{Rational Numbers} \label{RN}
          As is well known the rational numbers correspond to
          equivalence classes of ordered pairs  of integers. Usually the
          class is represented by the one ordered pair $\{p,q\}$
          where $p$ and $q$ are relatively prime and the rational
          number is represented in the form $p/q$.  Rational
          numbers are also axiomatizable by the field axioms.
          These are the axioms for a commutative ring with
          identity plus the axiom stating the existence of a
          multiplicative inverse \cite{Adamson}.

          The representation of rational numbers as pairs of
          integers has the disadvantage that the multiplication
          and especially the addition operations are rather
          opaque and unrelated to simple physical operations.
          Also they are not the representation used in
          computers that operate on single strings of symbols as
          rational approximations to real numbers.

          In particular, the sum of the two rational numbers $(a,b)=a/b$ and
          $(c,d)$  where $a,b,c,d$ are integers is the rational
          number $(a\times d + c\times b,b\times d)=[a\times d
          +c\times b]/[b\times d]$. Efficient implementation of this
          operation is possible, as it is based on efficient
          implementation of addition and multiplication of the
          integers.  However, the use of this fairly complex
          combination of integer addition and multiplication to
          represent a basic operation of addition of rational numbers,
          which is simple for the string representation, is one
          reason the integer pair representation is not used. Also
          the string representation is well suited to describe
          rational number approximations to real numbers.

          For these reasons the description of integers as tensor
          product states over the sites  $j=1,2,\cdots$ will be extended here to
          tensor product states over the sites
          $j = \cdots , -1,0,1,\cdots $. This description has the
          advantage that the basic successor and addition
          operations already defined can be easily adapted.  Also
          elementary multiplication operations corresponding to
          physical shifts are easy to define and are physically
          relatively easy to implement.

          This representation has the obvious disadvantage that
          many rational numbers as infinite repeating $"k-als"$
          are only approximately represented.  Only those rational
          numbers $p/q$ where all prime factors of $q$ are also
          factors of $k$ can be represented exactly as finite
           tensor product states.  In spite of this the
          importance of the requirement of efficient physical
          implementation and the fact that these are used in computations
          as rational approximations to real numbers outweighs the
          disadvantages.

          The corresponding tensor product states  in Fock space $|\pm\underline{r}\rangle$
          have the form \begin{eqnarray}|\pm\underline{r}\rangle
             = a^{\dagger}_{\pm ,n+1}a^{\dagger}_{
           \underline{r}(n),n}& \nonumber \\  \cdots a^{\dagger}_{
           \underline{r}(1),1}a^{\dagger}_{
           .,0}a^{\dagger}_{
           \underline{r}(-1),-1} & \cdots a^{\dagger}_{
           \underline{r}(-m),-m}|0\rangle.
           \label{rarep} \end{eqnarray}
           Here  $ \underline{r}$ is a
           function from the interval $[n,-m]$ to
           $0,1,\cdots ,k-1$ with $ \underline{r}(0)="."$, the "$k-al$" point.

           It is sometimes convenient to represent the state
           $|\pm\underline{r}\rangle$ as \begin{eqnarray}
           |\pm\underline{r}\rangle  =  |\pm\underline{s.t}\rangle
            = &  a^{\dagger}_{\pm ,L_{s}+1}a^{\dagger}_{
           \underline{s}(L_{s}),L_{s}}\cdots   \nonumber \\  a^{\dagger}_{
           \underline{s}(1),1}a^{\dagger}_{
           .,0}a^{\dagger}_{
           \underline{t}(-1),-1}& \cdots a^{\dagger}_{
           \underline{t}(-L_{t}),-L_{t}}|0\rangle. \label{rastrep}
           \end{eqnarray}
           Here $ \underline{s},\; \underline{t}$ are as defined
           before with $L_{s}$ and $L_{t}$ the lengths of $
           \underline{s}$ and $ \underline{t}$.

           As was the case for integers and natural numbers,
           states with leading or trailing strings of zeros will
           be excluded even though they represent the same
           rational number.  To this end the Hilbert space
           ${\mathcal H}^{Ra}$ of rational number states is the
           subspace of Fock space spanned by states of the form
           $|\pm\underline{s.t}\rangle$ where $ \underline{s}(L_{s})>0$ if
           $L_{s}>1$ and $ \underline{t}(-L_{t})>0$ if
           $L_{t}>1$. ${\mathcal H}^{Ra}$ also includes the state
           $|+\underline{0.0}\rangle =
           a^{\dagger}_{+,2}a^{\dagger}_{0,1}
           a^{\dagger}_{.,0}a^{\dagger}_{0,-1}|0\rangle$ which
           represents the number $0$.  Properties of operators for
           basic operations will be defined relative to this
           space.

           Here the component systems associated with each site
           are taken to be either bosons or fermions of the same type.
           Thus for each site $j$ the there must be $k+3$ states
           available to the boson or fermion as there are the
           states $|+,j\rangle ,\; |-,j\rangle ,\; |.,j\rangle$
           as well as the $k$ number states available to each
           system. If desired, one can construct a representation
           using fermions or bosons of different types for the
           sign  and $"k-al"$ point states.

           Also, as was the case for the natural numbers and integers,
           there are no problems here with the
           anticommutation relations for fermions provided the
           ordering shown in Eqs. \ref{rarep} and \ref{rastrep} is
           preserved.  The operators will be defined so that they
           do not generate any sign changes for fermion states.

           \subsection{The Successor Operators} \label{TSO}
           As was the case for the integers,  successor operators,
           $R_{j}$, can be defined for rational numbers. It is
           quite useful to follow the definition of $I_{j}$ and split the
           definition of $R_{j}$ into two cases:
           \begin{equation}
           R_{j}= R^{+}_{j}+R^{-}_{j}= R^{+}_{j}+ R^{-}_{\geq
           j}+R^{-}_{<j}.
            \label{Rdef} \end{equation} Here $R^{+}_{j}$ and
           $R^{-}_{j}$ act on the subspaces ${\mathcal H}^{Ra+}$ and
           ${\mathcal H}^{Ra-}$ corresponding to the subspaces of
           positive and negative rational number states
           respectively. For $j>0$  [$j<0$] these operators correspond informally to
           the addition of $k^{j-1}$  [$k^{j}$]. Also $R^{+}_{j}=R^{+}_{j}P_{+}$,
           $R^{-}_{\geq j}=R^{-}_{\geq j}P_{-}P_{\neq 0,\geq j}$,
           and $R^{-}_{<j}=R^{-}_{<j}P_{0,\geq j}$. The projection
           operators $P_{+}$ and $P_{-}$  are given by Eq.
           \ref{proj+-}, and
           \begin{eqnarray}
           P_{\neq 0,\geq j} & = &
           \sum_{\ell=j}^{\infty}\sum_{h=1}^{k-1}a^{\dagger}_{h,\ell}a_{h,\ell}
           \nonumber \\ P_{0,\geq j}& =&
           \sum_{\ell=j}^{\infty}a^{\dagger}_{0,\ell}a_{0,\ell}.
           \label{raproj} \end{eqnarray} These definitions are set
           up so that  $R^{-}_{\geq j}$ adds $k^{j}$ to negative numbers whose magnitude is
           $\geq k^{j}$, and $R^{-}_{<j}$ adds $k^{j}$ to negative numbers whose magnitude is
           $< k^{j}$.  In this last case the sign of the rational
           number is changed.

           Two cases need to be considered: $j>0$ and $j<0$.  For
           $j>0$ it is clear that \begin{eqnarray} R_{j} = I_{j}, &
           R^{+}_{j} = I^{+}_{j} \nonumber \\ R^{-}_{\geq j} = I^{-}_{\geq
           j}, & R^{-}_{<j} = I^{-}_{<j} \label{R=I>0} \end{eqnarray}
           The reason the definitions are the same for
           rational numbers and integers is
           that for $j>0$ the actions of $R_{j}$ are insensitive to the
           presence or absence of component systems at sites $<1$.

           For $j<0$, definitions of $R^{+}_{j},\;R^{-}_{\geq j},\; R^{-}_{<j}$
           can be given that are  similar to those given for the
           corresponding $I$ components. For  $R^{+}_{j}$ one has
           \begin{equation}
           R^{+}_{j}=\Gamma^{+}_{j}Y^{+}_{j} \label{defR+}
           \end{equation} where
           \begin{eqnarray}
           \Gamma^{+}_{j}=\sum_{h=0}^{k-2}a^{\dagger}_{h+1,j}a_{h,j}
           +\Gamma^{+}_{j+1}a^{\dagger}_{0,j}a_{k-1,j} \nonumber \\
           Y^{+}_{j}=P_{occ,j}+ Y^{+}_{j+1}a^{\dagger}_{0,j}P_{unnoc,j}.
           \label{defgammay}\end{eqnarray}  These equations are
           valid for $j\leq -2$ for $\Gamma^{+}_{j}$ and $Y^{+}_{j}$.
           For $j=-1$,  $Y^{+}_{-1}=1$ and  $\Gamma^{+}_{-1}$ is
           given by Eq. \ref{defgammay} with
           $K^{+}_{1}$ (Eq. \ref{K1succ}) replacing $\Gamma^{+}_{0}$ in the
           definition. $Y^{+}_{j}$ acts on only  those states
           $|+\underline{s}.\underline{t}\rangle$ where
           $j<-L_{t}-1$ by adding a string of $0s$ to the right,
           as in adding $10^{-7}$ to $63.04$. Otherwise it is the
           identity.

           The definition of $\Gamma^{+}_{j}$ is valid for both
           boson and fermion systems.  This follows from the fact
           that all terms contain an even number of annihilation
           and creation (a-c) operators with the result that anticommuting
           terms past other such operators to the point of action
           does not generate a sign change.  This is not the case
           for $Y^{+}_{j}$ for states in which this operator is
           active. These consist of states
           $|+\underline{s}.\underline{t}\rangle$ for which
           $|j|>L_{t}$. (Recall that $j<0$.) The problem here is
           that for many states
           $|+\underline{s}.\underline{t}\rangle$ moving $Y^{+}_{j}$
           to its point of action requires anticommuting an odd
           number of a-c operators past an odd number of a-c
           operators describing the state, giving a sign change.

           This can be avoided by redefining $Y^{+}_{j}$ for the
           fermion case to be
           \begin{eqnarray}
           Y^{+,f}_{j}& =
           P_{occ,j}-Y^{+,f}_{j+1}a^{\dagger}_{0,j}P_{unocc,j}(
           \sum_{\ell=2}^{\infty}(-1)^{\ell}P_{+,\ell}
           \nonumber \\& \times
           \sum_{m=-j}^{-1}(-1)^{m}P_{occ,m}P_{unocc,m-1}).
           \label{defyf} \end{eqnarray} where $P_{+,\ell}=a^{\dagger}_{+,\ell}
           a_{+,\ell}$. To see that sign changes
           are avoided one has $$\begin{array}{l}
           Y^{+,f}_{j}|+\underline{s}.\underline{t}\rangle =
           (-1)^{L_{s}+2+L_{t}}Y^{+,f}_{j+1}a^{\dagger}_{0,j}P_{unocc,j}
           |+\underline{s}.\underline{t}\rangle \\
           =(-1)^{2L_{s}+2L_{t}+4}Y^{+,f}_{j+1}+|\underline{s}.\underline{t}*0_{j}\rangle
           \end{array}$$ where
           $$\begin{array}{l}|+\underline{s}.\underline{t}*0_{j}\rangle=
           a^{\dagger}_{+,L_{s}+1}a^{\dagger}_{
           \underline{s}(L_{s}),L_{s}}\cdots a^{\dagger}_{
           \underline{s}(1),1} \\ \times a^{\dagger}_{.,0}\cdots
           a^{\dagger}_{
           \underline{t}(-L_{t}),-L_{t}}a^{\dagger}_{0,j}|0\rangle .\end{array}$$
           Since the exponent of $-1$ is even, this shows that no
           sign change occurs. Iterative application of
           $Y^{+,f}_{j+1}$, etc. causes no sign change because
           the added operators all stand to the left of
           $a^{\dagger}_{0,j}$ in order of increasing $j$.

           For $R^{-}_{\geq j}$ one has a result similar to Eq.
           \ref{I-geqj}: \begin{equation}
           R^{-}_{\geq j} =(P_{+0.0}+WP_{\neq 0,\geq
           j})(R^{+}_{j})^{\dagger}WP_{-}P_{\neq 0,\geq j}
           \label{R-geqj} \end{equation} where $P_{\neq 0,\geq
           j}=\sum_{\ell=j}^{\infty}\sum_{h=1}^{k-1}a^{\dagger}_{h,\ell}a_{h,\ell}$
           is the projection operator for finding a qubyte in
           state $|h,\ell\rangle$ with $h\neq 0$ and $\ell \geq
           j$. $P_{-} = \sum_{\ell
           =2}^{\infty}a^{\dagger}_{-,\ell}a_{-,\ell}$ is the
           projection operator on all negative rational number
           states.  $W$ is the sign change operator of Eq, \ref{sgnch}.
           This equation is based on the fact that  for
           all $j$ $(R^{+}_{j})^{\dagger}$ corresponds to
           subtraction of $k^{j-1}$ if $j>0$ and of $k^{j}$ if
           $j<0$ over its domain of definition. It expresses the
           correspondence $k^{j} -x.xxxxx=-(x.xxxxx-k^{j})$ for
           the case that $x.xxxxx\geq k^{j}$.

           For $R^{-}_{<j}$ for $j<0$ one has an equation similar to Eq.
           \ref{I-<j}: \begin{equation}
           R^{-}_{<j}= \sum_{ \underline{t}}(R^{+}_{
           \underline{t}})^{\dagger}R^{+}_{j}(R^{+}_{
           \underline{t}})^{\dagger}WP_{-\underline{t}}P_{0,\geq
           j}.
           \label{R-<j} \end{equation}
           This equation is based on the result that, as was the case for
           the integers, one sees that any rational number state
           $|+\underline{0}.\underline{t}\rangle$ can be written in
           the form \begin{equation}
           |+0.\underline{t}\rangle = R^{+}_{ \underline{t}}|+0.0\rangle
           \label{0-to-t+} \end{equation}  where \begin{equation}
           R^{+}_{ \underline{t}}=(R^{+}_{-1})^{
           \underline{t}(-1)}(R^{+}_{-2})^{
           \underline{t}(-2)}\cdots (R^{+}_{-L_{t}})^{
           \underline{t}(-L_{t})}. \label{R+tdef} \end{equation}

           This shows that for any state
           $|-0.\underline{t}\rangle$,
           $(R^{+}_{ \underline{t}})^{\dagger}W|-0.\underline{t}\rangle
           =|+0.\underline{0}\rangle$.  Application of
           $(R^{+}_{ \underline{t}})^{\dagger}R^{+}_{j}$ to this
           state gives the positive rational number state
           corresponding to the rational number $k^{j}-t$. This sequence of operations
           is expressed by Eq. \ref{R-<j}. The projection operator
           $P_{0,\geq j}$, in effect, limits the $ \underline{t}$
           sum to states for which $ \underline{t}(\ell)=0$ for $-1\geq \ell\geq j$.

           The operator $R_{j}$ has the same properties as $I_{j}$
           in that it is a bilateral shift,
           $(R_{j})^{\dagger}R_{j}=1=R_{j}(R_{j})^{\dagger}$ and
           \begin{equation} (R_{j})^{k} =\left \{
           \begin{array}{ll} R_{j+1} & \mbox{ if $j\neq -1$} \\
           R_{1} & \mbox{ if $j=-1$}\end{array}\right.
           \label{logeffR} \end{equation}
           For positive values of $j$ these results are immediate
           as $R_{j}=I_{j}$ and these properties have already been
           proved for $I_{j}$. For negative values of $j$ the
           proof should be essentially the same as that for the positive
           values of $j$ as the form and action of the operators $R^{-}_{\geq
           j},R^{-}_{<j}$ is essentially the same as that for the
           corrresponding integer operators.

           From Eq. \ref{logeffR} one has results similar to Eq.
           \ref{999I}: \begin{equation}
           (R_{j})^{\dagger}R_{j+n} = (R_{j})^{k-1}\cdots
           (R_{j+n-1})^{k-1}. \label{999Rss} \end{equation}
           This holds for all positive $n$ and all $j$ such that
           either $j$ and $j+n$ are both positive or both are
           negative.  In case $j$ is negative and $j+n$
           positive one has \begin{eqnarray}
           (R_{j})^{\dagger}R_{j+n}=(R_{j})^{k-1}\cdots
           (R_{-1})^{k-1} \nonumber \\ \times (R_{+1})^{k-1}\cdots
           (R_{j+n-1})^{k-1}. \label{999Rds} \end{eqnarray}  If
           $j,n$ are such that $j+n=1$ then the subscript $j+n-1$
           is replaced by $j+n-2$ in the above.

           \subsection{Rational Addition and Multiplication} \label{RAM}
           The definition of addition for rationals is quite
           similar to that for the integers. One has
            \begin{equation}
            \widetilde{+}_{R}|\pm\underline{p}\rangle\otimes
            |\pm\underline{q}\rangle = |\pm\underline{p}\rangle\otimes
            |\pm\underline{q}+\pm\underline{p}\rangle
            \label{defplusR}
            \end{equation}
            where $|\underline{p}\rangle ,|\underline{q}\rangle$
            have the form of $|\underline{r}\rangle$ of Eq.
            \ref{rastrep}. Also for $|\pm\underline{p}\rangle =
            |\pm\underline{s}.\underline{t}\rangle$
            \begin{eqnarray}
            |\pm\underline{q}+\pm\underline{p}\rangle &=&(R_{L_{s}})^{\pm
          \underline{s}(L_{s})}\cdots (R_{1})^{\pm
          \underline{s}(1)} \nonumber \\ &\mbox{} & (R_{-1})^{\pm
          \underline{t}(-1)}\cdots (R_{-L_{t}})^{\pm
          \underline{t}(-L_{t})} |\pm\underline{q}\rangle  \nonumber \\  &=&  R_{\pm
          \underline{p}}|\pm\underline{q}\rangle.
          \label{defplusRpm}
          \end{eqnarray}
          $ \widetilde{+}_{R}$ has the same properties as $
          \widetilde{+}_{I}$. It is unitary on ${\mathcal H}^{ra}$
          and the adjoint corresponds to the subtraction operator.

          The definition of multiplication has the same form as
          that for the integers in Eq. \ref{defItimes1}. However
          the definition of the shift operator $U^{R,i}$,
          corresponding to multiplication by $k$ is more complex.

          There are several ways to define $U^{R,i}$.  Here the
          operator $U^{R,i}$ corresponding to multiplication by $k$,
          acting on a state $|\pm \underline{p}\rangle$, first exchanges
          the point at site $0$ with the number at site $-1$.  Then the
          whole state is shifted one site to the left.  $0$ is added to
          site $-1$ if and only if the site becomes unoccupied.
          That is, \begin{equation}U^{R,i}|\pm\underline{s}.\underline{t}\rangle =
          |\pm\underline{s* \underline{t}(-1)}.\underline{t^{\prime}}\rangle
          \label{defUR}\end{equation} where $
          \underline{t^{\prime}}(j)=\underline{t}(j-1)$ for $-1\geq
          j\geq -L_{t}+1$ if $L_{t}>1$ and
          $\underline{t^{\prime}}(-1)=0$ if $L_{t}=1$.

          The definitions are the same for bosons and fermions
          except for the case when $0$ must be added.
          For bosons these operations are defined by
          \begin{equation}
          U^{R,b} =
          Z\sum_{h=0}^{k-1}a^{\dagger}_{h,0}a_{.,0}a^{\dagger}_{.,-1}a_{h,-1}
          \label{defRUsumb} \end{equation} where
          $Z=\sum_{j=2}^{\infty}Z_{j}$.  $Z_{j}$ is given by
          \begin{equation}
          Z_{j}= \left \{ \begin{array}{ll}
          Z_{j-1}\sum_{h}a^{\dagger}_{h,j+1}a_{h,j}P_{unocc,j+1} &
          \mbox{if $j\geq 0,\neq 2$} \\ \\ \begin{array}{l}
          (Z_{j-1}P_{occ,j-1}+ P_{unocc,j-1}) \times \\ \\ \sum_{h}
          a^{\dagger}_{h,j+1}a_{h,j}P_{unocc,j+1}\end{array}
          & \mbox{if $j\leq -2$} \end{array}\right. .
          \label{defZR>j} \end{equation}For $j=2$ \begin{eqnarray}
Z_{2} = Z_{1}\sum_{h}a^{\dagger}_{h,3}a_{h,2}(P_{numocc,2}+P_{\pm, 2}P_{\neq 0,1})
\nonumber \\
+ Z_{-1}\sum_{h=0}^{k-1}a^{\dagger}_{\pm ,2}a^{\dagger}_{h,1}a_{h,0}a_{\pm , 2}P_{\pm
,2}P_{0,1}. \label{defZ2} \end{eqnarray} For $j=-1$
          \begin{eqnarray}
          Z_{-1} = Z_{-2}\sum_{h}a^{\dagger}_{h,0}a_{h,-1}P_{unocc,0}P_{occ,-2}
          \nonumber \\
          +\sum_{h}a^{\dagger}_{h,0}a^{\dagger}_{0,-1}a_{h,-1}
          P_{unocc,0}P_{unocc,-2}. \label{defZR>-1} \end{eqnarray}
          Note that some of the $h$ sums over $0,\cdots k-1$ also include sums over
          $.,+,-,$.  The subscripts on the projection operators are self explanatory. $P_{numocc,2}$
is the projection operator for a qubyte state $|-,2\rangle$ where $-$ denotes a number in
$0,\cdots,k-1$.

To understand the reason for singling out $Z_{2}$ and $Z_{-1}$
one notes that the action of $U^{R,b}$ is given by $$U^{R,b}|\pm
\underline{s}.\underline{t}\rangle
=Z|\pm\underline{s}*\underline{t}(-1)_{0}._{-1}\underline{t}_{[-2,-L_{t}]}\rangle.
$$  The cases where $|\pm \underline{s}.\underline{t}\rangle =
|\pm 0.\underline{t}\rangle$ or $|\pm\underline{s}.0\rangle$ need
special treatment in order to comply with the convention that no
leading or trailing strings of $0s$ remain. For $|\pm
\underline{s}.\underline{t}\rangle = |\pm
0.\underline{t}\rangle$, $Z_{2}$ acts on
$|\pm_{2}0_{1}\underline{t}(-1)_{0}._{-1}\underline{t}_{[-2,-L_{t}]}\rangle$
to delete the $0_{1}$ component before the shifting.  For $|\pm
\underline{s}.\underline{t}\rangle = |\pm\underline{s}.0\rangle$,
$Z_{-1}$ acts on the shifted state $|\pm\underline{s}*0.\rangle$
to add a $0_{-1}$ component to give
$|\pm\underline{s}*0.0\rangle$ as the final result.

          For fermions one has \begin{eqnarray}
          U^{R,f} = U^{R,b}P_{occ,-2}+ \nonumber \\
       Z^{f}\sum_{h=0}^{k-1}a^{\dagger}_{h,0}a_{.,0}a^{\dagger}_{.,-1}a_{h,-1}P_{unocc,-2}
          \label{defRUsumf} \end{eqnarray} where
          $Z^{f}=\sum_{j=2}^{\infty}Z^{f}_{j}$.  $Z^{f}_{j}$ is given
          by \begin{equation}
          Z^{f}_{j}=
          -Z^{f}_{j-1}\sum_{h=0}^{k-1,\pm}a^{\dagger}_{h,j+1}a_{h,j}P_{unocc,j+1}
          \label{defZRf>j} \end{equation} if $j\geq 0$ and
          \begin{equation} Z^{f}_{-1}=
          -\sum_{h=0}^{k-1,.}a^{\dagger}_{h,0}a^{\dagger}_{0,-1}
          a_{h,-1}P_{unocc,0} \label{defZRf>-1} \end{equation} if $j=-1$.  The range
          of the $h$ sums is denoted by the superscripts shown in
          the above.

          $U^{R,f}$ is defined to have the same action on
          fermion states as $U^{R,b}$ does on boson states.
          The only case in which the definition of $U^{R,b}$ and $U^{R,f}$
          differ is for the action on
$|\pm\underline{s}.0\rangle$ when $Z_{-1}$ is finally active at
the end of the shifting. The definition is set up so that
anticommuting the odd number of a-c operators to the right hand
end of $|\pm\underline{s}*0.\rangle$ to add the $0$ does not
change the sign.  The case in which a $0$ is deleted causes no
problems because there is no anticommuting of an odd number of a-c
operators.

$U^{R,i}$ has the property that, for $i=b,f$,  it is a bilateral
shift on ${\mathcal H}^{ra}\ominus |0.0\rangle$.  It is clear
from the definition that for all
$|\pm\underline{s}.\underline{t}\rangle \neq |0.0\rangle$,
$\langle\pm\underline{s}.\underline{t}|U^{R,i}|\underline{s}
.\underline{t}\rangle=0$. Also
$(U^{R,i})^{\dagger}U^{R,i}=1=U^{R,i}(U^{R,i})^{\dagger}$ as
$U^{R,i}$ is a bijection on the basis
$\{|\pm\underline{s}.\underline{t}\rangle\}$ spanning ${\mathcal
H}^{ra}$.

It follows from the definition of  $U^{R,i}$ that $U^{R,i}$ or
$(U^{R,i})^{\dagger}$ correspond to multiplication by $k^{j-1}$ or $k^{-j}$ respectively.
Based on this the multiplication operator $\widetilde{\times}$ is defined by
          \begin{equation}
          \widetilde{\times}|\pm\underline{p}\rangle |\pm\underline{q}\rangle
|\pm\underline{r}\rangle
          =|\pm\underline{p}\rangle |\pm\underline{q}\rangle
          |\pm\underline{r}+(\pm\underline{p}\times
          \pm\underline{q})\rangle. \label{defRtimes}
          \end{equation} Here
$|\pm\underline{r}+(\pm\underline{p}\times\pm\underline{q})\rangle$
is the state denoting the result of adding $\pm p\times \pm q$ to
$\pm r$. Following Eq. \ref{defItimes1} for the
          integers and using $|\pm \underline{p}\rangle
          =|\pm \underline{s}.\underline{t}\rangle$ one can express $\widetilde{\times}$ more
explicitly as \begin{eqnarray}
          \widetilde{\times}|\pm\underline{s}.\underline{t}\rangle |\pm\underline{q}\rangle
|\pm\underline{r}\rangle =|\pm\underline{s}.\underline{t}\rangle
((U^{R,i}_{2})^{\dagger})^{L_{s}-1}\nonumber
\\ \times (\widetilde{+}_{2,3})^{\pm \underline{s}(L_{s})}U^{R,i}_{2}
           (\widetilde{+}_{2,3})^{\pm \underline{s}(L_{s}-1)}
          U^{R,i}_{2} \cdots \nonumber \\ \times U^{R,i}_{2}
          (\widetilde{+}_{2,3})^{\pm \underline{s}(1)}U^{R,i}_{2}
          (\widetilde{+}_{2,3})^{\pm \underline{t}(-1)}  \cdots \nonumber \\ \times U^{R,i}_{2}
(\widetilde{+}_{2,3})^{\pm \underline{t}(-L_{t}+1)}U^{R,i}_{2}
(\widetilde{+}_{2,3})^{\pm \underline{t}(-L_{t})}\nonumber
\\ \times ((U^{R,i}_{2})^{\dagger})^{L_{t}} |\pm \underline{q}\rangle|\pm
\underline{r}\rangle. \label{defRtimes1}
\end{eqnarray} These actions correspond to multiplying $\pm q$ by
$k^{-L_{t}}$,  adding or subtracting $\underline{t}(-L_{t}) $
copies of the result to the third state $\pm r$, then adding or
subtracting $\underline{t}-(L_{t}+1) $ copies of
$k^{-L_{t}+1}(\pm q)$ to the third state, etc.. The last step
recovers the original second state $|\pm\underline{q}\rangle$ by
multiplying by $k^{L_{s}-1}$. Whether $\widetilde{+}$ carries out
iterated addition or subtraction depends on the sign of the first
state.

It is clear from the definition that $\widetilde{\times}$ is
unitary. The operator preserves orthonormality of the basis set
$\{|\pm\underline{p},\pm\underline{q},\pm\underline{r}\rangle \}$
and all these states ( and linear superpositions) are in the
domain and range of the operator. As was the case for the
integers, it does not follow from unitarity that the adjoint of
$\widetilde{\times}$ carries out division. The argument is
similar to that given for  the absence of a division operator for
the integers in that an equation similar to Eq. \ref{defIdiv}
would have to be satisfied. The fact that this is not the case is
a consequence of the fact that not all rational numbers are
included in the representation used here.

\subsection{The Rational Number Axioms}
The axioms for rational numbers are those for a field
\cite{Adamson}.  These are the same as those for the integers
with the added axiom stating the existence of an inverse to
multiplication. However as was seen this is not valid for the
representation used here. The proofs of the other axioms are quite
similar to those for the integers and the natural numbers and will
not be repeated here. The main difference here between the
rational number and integer operators is that the operator
$U^{R,i}$ is unitary whereas the corresponding operator $U^{I,i}$
for integers, Eq. \ref{UIsum}, is not unitary.

\section{Physical Models of the Axiom Systems}
\label{PMAS} So far mathematical Hilbert space models have been
constructed for the natural number, integer, and rational number
axiom systems.  However these models are all abstract in that
nothing is implied about the existence of physical systems that
can implement the operations described by the axiom systems.  The
ubiquitous existence of computers shows that  such systems do
exist, at least on a macroscopic or classical scale.

Here the emphasis is on microscopic  quantum mechanical systems.
These systems have the property that the switching time $t_{sw}$
to carry out a single step is short compared to the decoherence
time $t_{dec}$, or $t_{sw}/t_{dec}\ll 1$ \cite{DiVincenzo}. For
macroscopic systems $t_{sw}/t_{dec}\gg 1$. The discussion will be
fairly brief and will be applied here to the rational number
system. Additional details for modular arithmetic on the natural
numbers are given elsewhere \cite{BenRNQM}.

Let $A,D$ be two sets of physical parameters for quantum systems.
For instance $A$ could be an infinite set of space positions and
$D$ a finite set of spin projections or excitation energies of
the systems.   The physical Fock space of states ${\mathcal
H}^{phy}$ for the system is spanned by states of the form
$c^{\dagger}_{d_{m},a_{m}}c^{\dagger}_{d_{m-1},a_{m-1}}\cdots
c^{\dagger}_{d_{1},a_{1}}|0\rangle$.  Here $m$ is an arbitrary
finite number,  $c^{\dagger}_{d_{j},a_{j}}$ is a creation operator
for a system with property  $d_{j},a_{j}$, where $a_{j}$ and
$d_{j}$ are values in $A$ and $D$ respectively. The operators
$c^{\dagger}_{d,a}$ and $c_{d^{\prime},a^{\prime}}$ satisfy
commutation or anticommutation relations similar to Eqs.
\ref{acom} and \ref{acom1} if the basic physical systems are
bosons or fermions.

Assume that the  basic mathematical and physical systems are both
either fermions or bosons. Then the a-c operators of both
${\mathcal H}^{phy-Ra}$ and ${\mathcal H}^{Ra}$ have the same
symmetry property. Let $W$ be an arbitrary isometry from the
abstract Hilbert space ${\mathcal H}^{Ra}$ to a subspace
${\mathcal H}^{phy-Ra}$ of ${\mathcal H}^{phy}$. Then $W$ and its
adjoint $W^{\dagger}$ restricted to ${\mathcal H}^{phy-Ra}$, are
unitary maps between ${\mathcal H}^{phy}$ and ${\mathcal
H}^{phy-Ra}$.

One can then define induced annihilation and creation operators
on ${\mathcal H}^{phy-Ra}$ according to \begin{equation}
a^{\dagger}_{W,h,j}=Wa^{\dagger}_{h,j}W^{\dagger}
\;\;\;a_{W,h,j}=Wa_{h,j}W^{\dagger}. \label{defaW}
\end{equation}   Here $j=1,2,\cdots$ and
$h=0,1,\cdots ,k-1$.

The corresponding rational number states on the physical state
space are given by
\begin{eqnarray}
W|\pm\underline{s}.\underline{t}\rangle =|\pm_{W}\underline{s}
_{W}._{W}\underline{t}_{W}\rangle \nonumber \\
=a^{\dagger}_{W,\pm,L_{s}+1}a^{\dagger}_{W,\underline{s}(L_{t}),L_{t}}\cdots
a^{\dagger}_{W,.,0} \nonumber \\
\times a^{\dagger}_{W,\underline{t}(-1),-1}\cdots
a^{\dagger}_{W,\underline{}(-L_{t}),-L_{t}}|0\rangle.
\label{defphystate} \end{eqnarray} The operators $R_{W,j},\;
\widetilde{+}_{W},\; \widetilde{\times}_{W}$ on the physical state
space that correspond to the  successor operators for each $j$
and the addition and multiplication operators on ${\mathcal
H}^{Ra}$ are given by the general relation for any operator $
\widetilde{O}$ on ${\mathcal H}^{Ra}$
$$\widetilde{O}_{W}=W\widetilde{O}W^{\dagger}.$$  Alternatively
the physical state space operators can be obtained by replacing
each creation and annihilation operator $a^{\dagger}_{\ell,j},\;
a_{\ell,j}$ by $a^{\dagger}_{W,\ell,j},\; a_{W,\ell,j}$ in the
definitions of the operators given in subsections \ref{TSO} and
\ref{RAM}.

As a very simple example of a map $W$ let $ \underline{g}$  and $
\underline{d}$ be one-one functions from the numbers $1,2,\cdots$
to $A$ and  from $\{0,1,\cdots k-1$ to $D$. Let $W$ be such that
$$a_{W,h,j}=c_{ \underline{d}(h),\underline{g}(j)},\;\;
a^{\dagger}_{W,h,j}=c^{\dagger}_{
\underline{d}(h),\underline{g}(j)}.$$  In this case the
elementary physical components  of a physical quantum system
correspond to the components of the abstract quantum system.  This
type of example was considered earlier for modular arithmetic on
the natural numbers \cite{BenRNQM}.

More complex examples in which $W$ maps the abstract components
onto collective degrees of freedom or multiparticle states can
also be constructed.  These types of examples give entangled
physical states similar to those considered in some quantum error
correction schemes \cite{LaFMi,DiVSh,RaHa,KnLaF,ViKnLaF} and in
decoherence free subspaces \cite{Lidar,Zanardi,Filippo}.
Topological and anyonic quantum states have also been considered
in the literature \cite{Lloyd,Freedman,Kitaev}.

These examples also illustrate the large number of possibilities
for constructing unitary maps from ${\mathcal H}^{Ra}$ to a
physical state space for quantum systems.  However it is too
general in the sense that an important restriction has been left
out.  In particular, as is well known, there are many physical
systems that are not suitable to represent or model mathematical
number systems. Such systems are also not useful as quantum
computers.

This feature has been realized for some time and several
approaches  have been discussed.  Requirements discussed in the
literature  for quantum computers include having well
characterized qubits, the ability to prepare a simple initial
state, the condition that $t_{dec}/t_{sw}\gg 1$, the presence of
unitary operators for a universal set of quantum gates or unitary
control of suitable subsystems, and the ability to measure
specific qubits or subsystem observables
\cite{ViKnLaF,DiV0002077}.

Here the condition is expressed by the requirement that the basic
operations described by the axioms of the system under
consideration must be efficiently implementable.  For the systems
studied here this means that the successor operations for each
$j$, $ \widetilde{+}$, and $ \widetilde{\times}$ must be
efficiently implementable.

This requirement means that for each of these operations there
must exist a unitary time dependent operator $U(t)$ in the
physical  model such that the action of $U(t)$ on suitable
physical system states corresponds to carrying out the operation.
This can be expressed more explicitly using the rational number
states and operators as an example.  For each state $|\pm
\underline{r}\rangle = a^{\dagger}_{\pm\underline{r}}|0\rangle$
let $P_{\pm\underline{r}}$ and $P_{ \widetilde{O},
\pm\underline{r}}$ be the projection operators on the states
$|\pm\underline{r}\rangle$ and $| \widetilde{O}
\pm\underline{r}\rangle$ where $ \widetilde{O}$ is any of the
successor operators, $R_{j},\: \widetilde{+}_{R},\;
\widetilde{\times}_{R}$ defined in Section \ref{RN}. Let
$P^{W}_{\pm\underline{r}}= WP_{\pm\underline{r}}W^{\dagger}$ and
$P^{W}_{ \widetilde{O}, \pm\underline{r}}=WP_{
\widetilde{O},\pm\underline{r}}W^{\dagger}$ be the corresponding
projection operators on the  physical states
$|\pm\underline{r}\rangle$ and $\widetilde{O}
|\pm\underline{r}\rangle$.  These operators are the identity on
all the environmental and ancillary degrees of freedom in the
overall physical system.

Let $\rho(0)$ denote the initial overall physical system density
operator  at time $0$.  Then
$P^{W}_{\pm\underline{r}}\rho(0)P^{W}_{\pm\underline{r}}=
\rho_{\pm\underline{r}}(0)$ is the initial physical system state
with the model subsystem in the state corresponding to
$|\pm\underline{r}\rangle$ under the map $W$. The time
development of $\rho_{\pm\underline{r}}(0)$ is given by some
unitary operator $U_{ \widetilde{O}}(t)$ with possible dependence
on $ \widetilde{O}$ indicated.  That is
$$\rho_{\pm\underline{r}}(t)= U_{ \widetilde{O}}(t)\rho_{\pm\underline{r}}(0)
U^{\dagger}_{ \widetilde{O}}(t).$$

Implementability of the operator $\widetilde{O}$ means that there
is a unitary evolution operator $U_{ \widetilde{O}}(t)$ and  an
initial system state $\rho(0)$ such that for each
$|\pm\underline{r}\rangle$ there is a time $t_{\pm\underline{r}}$
such that the components of
$\rho_{\pm\underline{r}}(t_{\pm\underline{r}})$ that correspond to
the state $ \widetilde{O} |\pm\underline{r}\rangle$ appear with
relative probability $1$.  That is \begin{equation} TrP^{W}_{
\widetilde{O},
\pm\underline{r}}\rho_{\pm\underline{r}}(t_{\pm\underline{r}}) =
Tr\rho_{\pm\underline{r}}(0) \label{Imp} \end{equation} where the
trace is taken over all degrees of freedom including ancillary
and environmental degrees that may be present.

Implementability also means that the operator $ U_{
\widetilde{O}}(t)$ must be physically implementable in that there
must exist a  physical procedure for implementing $U_{
\widetilde{O}}(t)$ that can actually be carried out.  For
Schr\"{o}dinger dynamics this means there must exist a
Hamiltonian $H_{ \widetilde{O}}$ that can be physically
implemented  such that $ U_{ \widetilde{O}}(t)=e^{-i H_{
\widetilde{O}}t}$.  Eq. \ref{Imp} must also be satisfied by $H_{
\widetilde{O}}$.

The requirement of efficiency means that for each state
$|\pm\underline{r}\rangle$ the time $t_{\pm\underline{r}}$
required to satisfy Eq. \ref{Imp} must be polynomial in the
length $L_{r}$  of $ \underline{r}$.  It cannot be exponential in
$L_{r}$.  The requirement also means that the space requirements
for physical implementation must also be polynomial in $L_{r}$.
If $H_{ \widetilde{O}}$ is implemented by circuits of quantum
gates, as in \cite{Beckman,Vedral}, then the number of gates in
the circuits must be polynomial in $L_{r}$.

Efficiency also means that the thermodynamic resources needed to
implement $H_{ \widetilde{O}}$ must be polynomial in $L_{r}$.
This places limitations on the value of $k$ in that for physical
systems occupying a given space-time volume it must be possible to
reliably distinguish $k$ alternatives in the volume \cite{Lloyd1}.

As was noted in the introduction, the efficiency requirement is
the reason that successor operators are defined separately for
each $j$ and the efficiency requirement is applied to each
operator.  If the requirement applied to just one of the
operators, and not to the others, then physical models could be
allowed in which $t_{\pm\underline{r}}$ would be exponential and
not polynomial in $ \underline{r}$ for these operators. This
follows from the exponential dependence of the $R_{j}$ on $j$ as
shown in Eq. \ref{logeffR}\footnote{It would be expected that
these models would be excluded by the efficiency requirement
applied to $ \widetilde{+}$ as this operator is defined in terms
of iterations of the different $R_{j}$.}. The fact that efficient
computation based on efficient implementation of the basic
arithmetic operations is so ubiquitous shows that there are many
methods of implementing the $R_{j}$ efficiently in classical
computers at least.  However this does not reduce the importance
of the efficiency requirement for these operators.

The existence of the Hamiltonians  $H_{ \widetilde{O}}$ that can
be carried out is in general a nontrivial problem.  For modular
arithmetic on the natural numbers the existence of quantum
circuits for the basic arithmetic operations $+$ and $\times$
\cite{Beckman,Vedral} suggests that such Hamiltonians may exist
for the basic arithmetic operations on distinguishable qubits.
However most physical models described to implement simple quantum
computations are based on a time dependent Hamiltonian that
implements a product of different unitary operators. In many of
these models the computation is driven by a sequence of
individually prepared laser pulses to carry out specified
operations. The possibility of describing this with a time
independent Hamiltonian that can be physically implemented  on
multiqubit systems is a question for the future.

Here the problem is more complex in that bosonic or fermionic
quantum computation methods would be needed on states with an
indeterminate number of degrees of freedom.  Work on this problem
for binary  fermions using the representation of a-c operators as
Pauli products of the standard spin operators \cite{Ortiz} is a
possible avenue but more needs to be done. For computations in an
interactive environment one may hope that the use of decoherence
free subspaces \cite{Lidar,Zanardi,Filippo}, stabilizer codes
\cite{Bravyi,Kitaev}, or other methods of error protection
\cite{LaFMi,DiVSh,RaHa,KnLaF} will be workable.

          \appendix
          \section{Appendix}
          \subsection{Proof of Some Natural Number Axioms}
          It is sufficient for the proof of $ \widetilde{\times}
          (V_{j}\otimes \widetilde{1}\otimes \widetilde{1})|\underline{s}\rangle\otimes
          |\underline{t}\rangle\otimes|\underline{x}\rangle =|\underline{s}\rangle\otimes
          |\underline{t}\rangle\otimes|\underline{y}\rangle$ where
          $|\underline{y}\rangle
          =|\underline{x}+\underline{s}\times
          \underline{t}+\underline{k^{j-1}t}\rangle$ to set
          $|\underline{x}\rangle=0$. An expression for the product
          is needed in which all powers of each $V_{j}$ are
          collected together. Repeating
          Eq. \ref{deftimes1}  for $|\underline{x}\rangle =|\underline{0}\rangle$ gives
          $$\begin{array}{c}\widetilde{\times}|\underline{s},\underline{t},\underline{0}\rangle
          =((U^{i}_{2})^{\dagger})^{L_{s}-1}(\widetilde{+}_{2,3})^{
\underline{s}(L_{s})}U^{i}_{2}
          (\widetilde{+}_{2,3})^{ \underline{s}(L_{s}-1)}
          U^{i}_{2}  \\ \cdots U^{i}_{2}
          (\widetilde{+}_{2,3})^{ \underline{s}(1)}
          |\underline{s},\underline{t},\underline{0}\rangle.
          \end{array}$$

          Use of Eqs. \ref{defplus}, \ref{0-to-s} and
          \ref{defshift} gives
          $$\begin{array}{l}\widetilde{\times}|\underline{s},\underline{t},\underline{0}\rangle
           = |\underline{s},\underline{t}\rangle \otimes
          (V_{L_{t}+L_{s}-1})^{ \underline{s}(L_{s})
          \underline{k^{L_{s}-1}t}(L_{t}+L_{s}-1)} \cdots \\
          (V_{L_{s}})^{ \underline{s}(L_{s})
          \underline{k^{L_{s}-1}t}(L_{s})}
          (V_{L_{t}+L_{s}-2})^{ \underline{s}(L_{s}-1)
          \underline{k^{L_{s}-2}t}(L_{t}+L_{s}-2)} \cdots \\
          (V_{L_{s}-1})^{ \underline{s}(L_{s}-1)
          \underline{k^{L_{s}-2}t}(L_{s}-1)}\cdots ,\cdots
          (V_{L_{t}+1})^{ \underline{s}(2)
          \underline{kt}(L_{t}+1)} \\ \cdots (V_{2})^{
          \underline{s}(2) \underline{kt}(2)}(V_{L_{t}})^{
          \underline{s}(1) \underline{t}(L_{t})}\cdots (V_{1})^{
          \underline{s}(1) \underline{t}(1)}|0\rangle
          \end{array}$$
          where use was made of $(U^{i})^{j} |\underline{t}\rangle
          =|\underline{k^{j}t}\rangle$ and if $
          \underline{k^{j}t}(n)=0$, then $V^{
          \underline{k^{j}t}(n)}=\widetilde{1}$.  These identity
          factors have been deleted in the above. Also $L_{s}$ and
          $L_{t}$ are the lengths of $ \underline{s}$ and $
          \underline{t}$.

          One now collects together all $Vs$ with the same
          subscript value. As noted before this commuting of the
          $Vs$ past one another causes no problems for either
          fermions or bosons.  There are two cases to consider $L_{t}\geq L_{s}$ and
          $L-{t}\leq L_{s}$ which differ only in index labeling. Carrying
          this out for $L_{t}\geq L_{s}$ and putting the
          $Vs$ in order of decreasing subscript values from left
          to right gives $$\begin{array}{l}(V_{L_{t}+L_{s}-1})^{ \underline{s}(L_{s})
          \underline{k^{L_{s}-1}t}(L_{t}+L_{s}-1)} \\ (V_{L_{t}+L_{s}-2})^{
\underline{s}(L_{s})
          \underline{k^{L_{s}-1}t}(L_{t}+L_{s}-2)+\underline{s}(L_{s}-1)
          \underline{k^{L_{s}-2}t}(L_{t}+L_{s}-2)} \\
          \cdots ,\cdots (V_{2})^{\underline{s}(2)
          \underline{kt}(2)+\underline{s}(1)
          \underline{t}(2)}(V_{1})^{ \underline{s}(1)
          \underline{t}(1)}|0\rangle.\end{array}$$

          A more explicit expression including the terms represented by
          the $\cdots$ is
          \begin{eqnarray}(V_{L_{t}+L_{s}-1})^{E_{0}}\cdots
          (V_{L_{t}+1})^{E_{L_{s}-2}} & \cdots
          (V_{m})^{G_{L_{s},m}}\nonumber \\
          \cdots (V_{L_{s}-1})^{F_{L_{s}-1}} \cdots &
          (V_{1})^{F_{1}}|0\rangle. \label{stimest}
          \end{eqnarray}
          Here \begin{equation} E_{n} =\sum_{h=0}^{n}
          \underline{s}(L_{s}-h) \underline{t}(L_{t}-n+h) \label{En} \end{equation}
          for $ 0\leq n\leq L_{s}-2$,\begin{equation}
          G_{L_{s},m}=\sum_{h=0}^{L_{s}-1} \underline{s}(L_{s}-h)
          \underline{t}(m+1+h-L_{s}) \label{Gst} \end{equation} for $ L_{s}\leq
          m\leq L_{t}$, and \begin{equation}  F_{\ell}=\sum_{h=0}^{\ell -1}
          \underline{s}(\ell-h) \underline{t}(h+1) \label{Fell} \end{equation} for
          $ 1\geq \ell \geq L_{s}.$   Note that
          $F_{L_{s}}=G_{L_{s},L_{s}}$ and
          $G_{L_{s},L_{t}}=E_{L_{s}-1}$.  Also
          \begin{equation}
          \underline{k^{n}t}(m+n)= \underline{t}(m) \label{knt}
          \end{equation} for
          $n=0,1,\cdots ,\; m=1,2,\cdots$ was used.

          In the above the values of the exponents may well be
          greater than $k-1$.  Because of Eq. \ref{logeff} this
          causes no problems provided one represents states in the
          form given by Eq. \ref{0-to-s}.

          The desired goal is to prove that $|(
          \underline{s+k^{j-1}})\times \underline{t}\rangle
          =|\underline{s\times t}+\underline{k^{j-1}t}\rangle$.

          From the forgoing one has
          \begin{eqnarray}
          |\underline{s\times t}+\underline{k^{j-1}t}\rangle & = &(V_{L_{t}+L_{s}-1})^{E_{0}}
          \nonumber \\ & \cdots &
          (V_{L_{t}+1})^{E_{L_{s}-2}}  \cdots
          (V_{m})^{G_{L_{s},m}}\nonumber \\
          & \cdots & (V_{L_{s}-1})^{F_{L_{s}-1}} \cdots
          (V_{1})^{F_{1}} \nonumber \\ (V_{L_{t}+j-1})^{ \underline{t}(L_{t})} &
          \cdots &
          (V_{j})^{ \underline{t}(1)}|0\rangle \label{stimest+kjt}
          \end{eqnarray}
          where Eq. \ref{knt} was used. Again one collects $Vs$ with
          the same subscripts.  The explicit form of the final result
          depends somewhat on the magnitude of $j$ relative to $L_{t}$
          and $L_{s}$.  Assume $j<L_{s}<L_{t}$.  Then the
          righthand side of Eq. \ref{stimest+kjt} can be written as
          $$\begin{array}{l}(V_{L_{t}+L_{s}-1})^{E_{0}}\cdots
          (V_{L_{t}+j})^{E_{L_{s}-j-1}}
          (V_{L_{t}+j-1})^{E_{L_{s}-j}+\underline{t}(L_{t})} \\ \cdots

(V_{L_{t}+1})^{E_{L_{s}-2}+\underline{t}(L_{t}-j+2)}(V_{L_{t}})^{G_{L_{s},L_{t}}
          +\underline{t}(L_{t}-j+1)} \\ \cdots (V_{m})^{G_{L_{s},m}
          +\underline{t}(m-j+1)}\cdots (V_{L_{s}})^{G_{L_{s},L_{s}}
          +\underline{t}(L_{s}-j+1)} \\ (V_{L_{s}-1})^{F_{L_{s}-1}
          +\underline{t}(L_{s}-j)}  \cdots (V_{j})^{F_{j}
          +\underline{t}(1)}\cdots (V_{1})^{F_{1}}|0\rangle.
          \end{array}$$

          The $E$ exponents containing $ \underline{t}$ can be
          written as $$\begin{array}{l} E_{L_{s}-j+p}+\underline{t}(L_{t}-p)= \\ \\
          \sum_{h=0}^{L_{s}-j+p} \underline{s}(L_{s}-h)
          \underline{t}(L_{t}-L_{s}+j-p+h)
          +\underline{t}(L_{t}-p)\end{array}$$ with $0 \leq p\leq j-2$.
          Similarly for the $G$ and $F$ exponents, $$\begin{array}{l}
          G_{L_{s},m}+\underline{t}(m-j+1)= \\ \\
          \sum_{h=0}^{L_{s}-1} \underline{s}(L_{s}-h)
          \underline{t}(m+1-L_{s}+h)
          +\underline{t}(m-j+1) \\ \\ F_{L_{s}-q}+\underline{t}(L_{s}-q-j+1)= \\ \\
          \sum_{h=0}^{L_{s}-q-1} \underline{s}(L_{s}-q-h)
          \underline{t}(h+1)
          +\underline{t}(L_{s}-q-j+1).\end{array}$$ Here $L_{s}\geq m\geq L_{t}$
          and $1\geq q\leq L_{s}-j$.

          These expressions can be rewritten as \begin{eqnarray}
          E_{L_{s}-j+p}+\underline{t}(L_{t}-p) = \nonumber \\
          \sum_{\begin{array}{c}
          {\scriptstyle h=0 }\\ {\scriptstyle h\neq L_{s}-j}\end{array}}^{L_{s}-j+p}
           \underline{s}(L_{s}-h)
          \underline{t}(L_{t}-L_{s}+j-p+h)+ \nonumber \\
          \mbox{} [\underline{s}(j)
          \underline{t}(L_{t}-p)+\underline{t}(L_{t}-p)]
          \label{E[stj]} \end{eqnarray}
          \begin{eqnarray}
          G_{L_{s},m}+\underline{t}(m-j+1) = \nonumber \\
          \sum_{\begin{array}{c}
          {\scriptstyle h=0 }\\ {\scriptstyle h\neq L_{s}-j}\end{array}}^{L_{s}-1}
           \underline{s}(L_{s}-h)
          \underline{t}(m+1-L_{s}+h)+ \nonumber \\
          \mbox{} [\underline{s}(j)
          \underline{t}(m-j+1)+\underline{t}(m-j+1)]
          \label{G[stj]} \end{eqnarray}
          \begin{eqnarray}
          F_{L_{s}-q}+\underline{t}(L_{s}-q-j+1) = \nonumber \\
          \sum_{\begin{array}{c}
          {\scriptstyle h=0 }\\ {\scriptstyle h\neq L_{s}-q-j}\end{array}}^{L_{s}-q-j}
           \underline{s}(L_{s}-q-h)
          \underline{t}(h+1)+ \nonumber \\
          \mbox{} [\underline{s}(j)
          \underline{t}(L_{s}-q-j+1)+\underline{t}(L_{s}-q-j+1)].
          \label{F[stj]} \end{eqnarray}

          Use of the axiom that is being proven for the operators
          $V_{j}, \widetilde{+},\widetilde{\times}$ for the number
          expressions in the square brackets in the above three
          equations gives
          \begin{equation} \underline{s}(j) \underline{t}(r) +
          \underline{t}(r) = ( \underline{s}(j)+1)
          \underline{t}(r). \label{str} \end{equation}
          for $r=L_{t}-p,m-j+1,L_{s}-q-j+1$.  This is the
          important step because by repeating the above derivation
          for $ \widetilde{\times}(V_{j}\otimes
          \widetilde{1}\otimes
          \widetilde{1})|\underline{s},\underline{t}\underline{0}\rangle$
          one can show that $|(V_{j} \underline{s})\times
          \underline{t}\rangle$ has exactly the form given by Eqs.
          \ref{E[stj]},\ref{G[stj]}, \ref{F[stj]} with $
          \underline{s}(j) \underline{t}(r)+\underline{t}(r)$
          replaced by $( \underline{s}(j)+1) \underline{t}(r)$.
          This completes the proof of the axiom for
          $j<L_{s}<L_{t}$.  The proof for the other cases is quite
          similar and will not be repeated here.

          \subsection{Proof of $I_{j+1} = (I_{j})^{k}$}

           For the proof of Eq. \ref{logeffI} one notes that  \begin{eqnarray} I^{k}_{j} &=&
          (I^{+}_{j})^{k}+  \sum_{\ell =1}^{k-1}[(I^{+}_{j})^{\ell
          -1}(I^{+}_{j}P_{0,+}W+I^{-}_{<j}P_{-,<
          j})\nonumber \\  \mbox{} & \times & (W(I^{+}_{j})^{\dagger}W)^{k-\ell}P_{-,\geq j}]
          \nonumber \\ \mbox{} &+&
(P_{0,+}W+P_{-,<0})(W(I^{+}_{j})^{\dagger}W)^{k}P_{-,\geq  j}
           \nonumber \\ \mbox{} & +&(I^{+}_{j})^{k-1}I^{-}_{<j}P_{<j}\nonumber.
          \end{eqnarray}
          The various terms  reflect the fact that $I^{-}_{<j}$ is
          either not active or is active just once (as it takes negative number states to
          positive number states). The $\ell$ sum shows that $I^{-}_{<j}$ can be active at any
          iteration of $I_{j}$, from the first to the $kth$.  It
          is preceded by iterations of $I^{-}_{\geq j}$ and
          succeeded by iterations of $I^{+}_{j}$.

          For the first term the result is immediate by Eq.
          \ref{logeffI+}.  Using $W^{2}=1$, the term
$(P_{+,0}W+P_{-,<0})(W(I^{+}_{j})^{\dagger}W)^{k}P_{-,\geq
          j}= (P_{+,0}W+P_{-})(W(I^{+}_{j+1})^{\dagger}W)P_{-,\geq
          j}$.  It is clear that this term gives $0$ for any state
          $|-\underline{s}\rangle$ for which $L=j$.  Thus
          $P_{-,\geq j}$ can be replaced by $P_{-,\geq j+1}$.
          This shows that this term equals $I^{-}_{\geq j+1}$.

          It remains to show that the sum of the remaining terms
          equals $I^{-}_{<j+1}$. For the last term, use of Eq.
          \ref{I-<j} gives $(I^{+}_{j})^{k-1}I^{-}_{<j}P_{<j}= (I^{+}_{j})^{k-1}\sum_{
          \underline{s}}^{L<j}(I^{+}_{\underline{s}})^{\dagger}I^{+}_{j}
          |+0\rangle\langle -,\underline{s}|$.  Commuting\footnote{That the operators
          $I^{+}_{j}$ and $I^{+}_{h}$ for $h\neq j$ and their adjoints do not,
          in general, commute
          or anticommute is clear from their definitions and the commutation
          relations for the creation and annihilation operators. However when applied
          to specific states of interest here, one sees that the results of the
          application is independent of the order in which they are applied.  Also
          there is no sign problem for fermions, provided the specific ordering
          of the component states described earlier is adhered to.} the
          $I^{+}_{j}$ past $(I^{+}_{\underline{s}})^{\dagger}$ and
          use of Eq. \ref{logeffI+} gives $\sum_{\underline{s}}^{L<j}
          (I^{+}_{\underline{s}})^{\dagger}I^{+}_{j+1}
          |+0\rangle\langle -,\underline{s}|$ for the last term.
          This corresponds in the $ \underline{s}$ sum for
          $I^{-}_{<j+1}$ to those terms for which $L<j$.

          The terms for which $L=j$ are contained in the $\ell$
          sum.  Each term can be written as
          $(I^{+}_{j})^{\ell}P_{+,0}W+(I^{+}_{j})^{\ell -1}\sum_{
          \underline{t}}^{L<j}(I^{+}_{\underline{t}})^{\dagger}I^{+}_{j}
          |+0\rangle\langle
          -,\underline{t}|)W(I^{+}_{j})^{\dagger})^{k-\ell}WP_{-,\geq
          j}$.  The matrix element $\langle
          -,\underline{t}|)W(I^{+}_{j})^{\dagger})^{k-\ell}W|-\underline{s}\rangle$
          is nonzero if and only if  the length $L$ of $
          \underline{s}$ is  $j$ and  $|- \underline{s}\rangle =
          a^{\dagger}_{-,j+1}a^{\dagger}_{k-\ell ,j}a^{\dagger}_{\underline{t}}|0\rangle$.
          In this case the matrix element equals $1$.  As a result the projection
          operator $P_{-,\geq j}$ can be replaced by $P_{-,j}$.

          Using this  and replacing $\ell$ by $k-\underline{s}(j)$ gives for
          the second part of the $\ell$ term $\sum_{
\underline{t}}^{L<j}(I^{+}_{j})^{k-\underline{s}(j)-1}
(I^{+}_{\underline{t}})^{\dagger}I^{+}_{j}
          |+0\rangle\langle
          -,\underline{t}|)W(I^{+}_{j})^{\dagger})^{ \underline{s}(j)}W|-
          \underline{s}\rangle\langle \underline{s}-|P_{-,j}=(I^{+}_{
           \underline{s}})^{\dagger}I^{+}_{j+1}|+0\rangle\langle-\underline{s}|P_{-,j}$.
           Here use is made of the facts that $|+ \underline{s}\rangle =
            |+\underline{s}(j)*\underline{t}\rangle$ and  $(I^{+}_{ \underline{t}})^{\dagger}
            ((I^{+}_{j})^{\dagger})^{ \underline{s}(j)} = (I^{+}_{
            \underline{s}})^{\dagger}$. Also
            $(I^{+}_{j})^{k-1}(I^{+}_{
            \underline{t}})^{\dagger}I^{+}_{j}|0\rangle =(I^{+}_{
            \underline{t}})^{\dagger}(I^{+}_{j})^{k}|0\rangle$ was
            used\footnotemark[2].

            The definition of $I^{-}_{<j}$ is such that the $
            \underline{t}$ sum is over all $ \underline{t}$ such
            that $\underline{t}(L)>0$. Thus states
            $|-\underline{s}\rangle = |-\underline{s}(j)*
            \underline{0}_{[j-1,1]}\rangle$ are excluded.  These
            are accounted for in the first part of the $\ell$ term:
            $(I^{+}_{j})^{\ell}P_{+,0}((I^{+}_{j})^{\dagger})^{k-\ell}WP_{-,\geq
            j}$.  Use of $ \underline{s}(j)=k-\ell$ and
            $(I^{+}_{j})^{\ell}=(I^{+}_{j})^{k-\underline{s}(j)}=(I^{+}_{
            \underline{s}})^{\dagger}I^{+}_{j+1}$ completes the
            proof of Eq. \ref{logeffI}.

          \end{document}